\documentclass[12pt,leqno]{article}
\usepackage{verbatim,color,amsmath,amssymb,graphicx,caption,ulem,booktabs}
\usepackage{multirow}
\usepackage{multicol}
\usepackage{amsmath}
\usepackage{amssymb}

\newcommand{\mbf}[1]{\mbox{\boldmath $#1$}}
\newcommand{\ba}{{\mbf \beta}}

{\catcode `\@=11 \global\let\AddToReset=\@addtoreset}
\AddToReset{equation}{section}

\AddToReset{Theorem}{section}

\newcommand{\cF}{{\cal F}}
\newcommand{\cG}{{\cal G}}

\def\ba{\begin{array}}
	\def\bc{\begin{center}}
		\def\bd{\begin{description}}
			\def\be{\begin{enumerate}}
				\def\ea{\end{array}}
			\def\ec{\end{center}}
		\def\ed{\end{description}}
	\def\ee{\end{enumerate}}
\def\ben{\begin{equation}}
	\def\benn{\begin{equation*}}
		\def\een{\end{equation}}
	\def\eenn{\end{equation*}}
\def\benr{\begin{eqnarray}}
	\def\eenr{\end{eqnarray}}
\def\benrr{\begin{eqnarray*}}
	\def\eenrr{\end{eqnarray*}}

\def\al{\alpha}
\def\b{\beta}

\def\del{\delta}

\def\g{\gamma}
\def\G{\Gamma}
\def\h{\hat}

\def\la{\lambda}
\def\lel{\label}

\def\mb{\mbox}
\def\noi{\noindent}

\def\nn{\nonumber}

\def\r{\ref}

\def\ro{\rho}

\def\si{\sigma}
\def\st{\sum_{t=1}^n}

\def\th{\theta}

\def\vep{\varepsilon}

\def\vs{\vskip}

\begin{document}

\def\Var{\mb{Var}}
\def\Cov{\mb{Cov}}
\def\so{\sout}
\def\tecr{\textcolor{red}}
\def\blu{\textcolor{blue}}
\def\grn{\textcolor{green}}
\def\red{\tecr}
\def\exp{\mb{exp}}
%

\bc
{\bf Zero-modified Count Time Series with Markovian Intensities}
\vs .25in

{ N. Balakrishna, Cochin University of Science and Technology, Kochi, India,\\ \quad P. Muhammed Anvar, Governmet Victoria College, Palakkad, India, \quad \\  Bovas Abraham, University of Waterloo, Canada.}\\

\ec
\vs .25in
\begin{abstract}
This paper proposes a method for analyzing count time series with inflation or deflation of zeros.  
 In particular, zero-modified  Poisson and zero-modified negative binomial series with intensities generated by non-negative Markov sequences are studied in detail. Parameters of the model are estimated by the method of estimating equations which is facilitated by expressing the model in a generalized state space form. The latent intensities required for estimation are extracted using generalized Kalman filter.  The applications of proposed model and its estimation methods are illustrated using simulated and  real data sets.  
\end{abstract}
MSC codes: Primary: 62M10; Secondary : 97K60.\\
Keywords: {\it{Estimating function; Generalized Kalman filter; Parameter driven models; Stochastic conditional duration; Zero inflation; Zero deflation}}
\section{Introduction}
Time series data in certain areas such as public health and environment are available often in the form of counts. For instance, weekly or monthly occurrence of certain disease in a city over time form a time series of counts. In air pollution studies, modelling annoyance caused by particulate matter such as dust and smoke (or by odor and noise), the observations often are counts and are dependent over time leading to the possible use of count time series models. In such situations it is natural to adopt a Poisson process or some other discrete time counting process to model the data.  Some of these time series such as monthly counts of workplace injuries or crimes in a region may contain large number of zeros which requires the use of what is known as zero inflation models or hurdle models. Another special case  found in practice is the case of smaller number of occurrences of zeros known as zero deflated data. Moreover, some of the count data may not have zeros. Such data may not be modelled using usual Poisson models. In such situations, zero modified models provide a more general frame work to take care of the count data with inflation or deflation or truncation at zero, when no information about the nature  of this situation is known. 

  A recent Handbook of discrete-valued time series edited by Davis et al. (2016) contains several theoretical developments and application avenues of count time series models.  Following Cox (1981) one can classify count time series models as (i) observation driven models or (ii) parameter driven models. In these count models, the mean and variance may depend on previous measurements and so it is natural to consider generalized autoregressive conditional heteroscedastic (GARCH) like models and refer to them as observation driven. On the other hand we can think of the mean and variance to have latent models as in Stochastic Volatility (SV) models and such models are called parameter driven. 

 Also, in modelling the number of transactions in financial markets, count series models with dependent intervals between transactions is a natural choice. In this context, Engle and Russell (1998) introduced the Autoregressive Conditional Duration (ACD) model. A simple ACD model can be described as follows: Suppose $X_t$ is a random variable (rv) denoting the duration between the ‘arrival times’ or times of transactions, then  
 $E(X_t|x_{t-1},x_{t-2}, ...)$ = $\la_{t}(x_{t-1},...,\th)$ or  $X_t = \la_t\vep_t,$  where $\{\vep_t\}$ is independent and identically distributed (iid) with density function $f(\vep_t,\boldsymbol{\beta}),$
$E(\vep_t) = 1,$ $\la_{t}$ is a function of the previous durations and parameters $\th,$  where as $\boldsymbol{\b}$ is a vector of parameters of the distribution with probability density function (pdf) $f$. 
This model is an example of an ‘observation driven model’ and can have many variations by specifying $\la_t$ and $f$ differently. It can also be generalised to address many different situations.  A recent survey on these models may be found in Bhogal and Variyam (2019).

The  method of construction of such models was extended to incorporate the count series $\{Y_t\}$ with observation driven intensities.   One such model is the observation driven model introduced in Ferland et al. (2006). A very simple case of this is given by the following. Conditional on the past, the counts, $Y_t,$ at time $t$,  has Poisson distribution,  $Y_t|\cF_{t-1} \sim P(\la_t)$, where $P$ is a Poisson distribution with intensity parameter $\la_t$  and $\cF_{t-1}$ contains all the information up to time $t-1.$ In addition, $\la_{t}$ depends on the previous observations through the model, $E(Y_t| y_{t-1},…)=\la_{t}=\gamma+\al y_{t-1}+ \b \la_{t-1}$ . One can see that this is a special case of the ACD model. 


  Tjostheim (2016) discussed the details of count time series under observation driven setup. 
 This model was generalised to the case of random coefficient models (see for instance Fokianos (2016)), where  $Y_t|\cF_{t-1} \sim P(Z_t\la_t)$, where $\{Z_t\}$ is an iid sequence of positive random variables with mean 1 and independent of $Y_i, i<t$. This is a case of count process with parameter driven intensity processes.  {Recently, Yang et al. (2015), proposed a flexible class of dynamic models for zero inflated count time series in the state-space framework and performed a sequential Monte Carlo  analysis.} 
 A key property of the class of parameter driven count series is that, although the observations are correlated marginally, they are independent, conditional on the latent process.

In this paper we study the properties of the count processes with modified frequency of  zeroes when their intensities are generated by certain latent parameter driven models. Naturally the proposed model is more general than zero inflated models. In the context of regression, Lambart (1992) proposed zero-inflated Poisson regression, to model the defects in manufacturing. Dietz and Bohning (2000) introduced the zero modified Poisson regression. Barreto-Souza (2015) studied the zero modified geometric INAR(1) model to analyze count time series with deflation or inflation of zeros. Sharafi et al. (2021) constructed a  first order integer valued autoregressive process with zero modified Poisson-Lindley distributed innovations to model zero modified count time  series.

To the best of our knowledge, the case of zero modified dynamic count time series is not discussed anywhere in the literature. We try to fill this gap. In particular we consider the zero modified count processes  when the intensities are generated by stationary Markov sequences. We list some specific examples of such Markov sequences in Section \r{sec2.3}. Statistical inference for observation-driven count processes are more easier to handle via likelihood based methods. The involvement of unobserved latent intensities make the inference difficult for the parameter driven models.

Estimating functions (EFs) are widely used in situations where the explicit form of the likelihood function is not available or intractable (Godambe, 1985). Naik-Nimbalkar
and Rajarshi (1995) have used this method in the context of state-space (SS) models, whereas Thavaneswaran et al. (2015) and
Thekke et al. (2016) have applied this method in the context of stochastic conditional duration models.

  We propose methods based on EF to estimate parameters, which are more straight forward compared to the MCMC methods. However, the resulting estimating equations depend on the latent intensities, which are not observable. To circumvent this problem, we propose to filter the intensities from the observed count data. In order to achieve that, we represent the model in a generalized state space (GSS) form and then adopt the generalized Kalman filter (GKF) algorithm proposed by Zenwirth (1988). Rest of the paper is organised as follows. In the next section, we introduce the zero modified count processes induced by latent Markov sequences referred to as zero modified stochastic conditional duration (ZMSCD) models. The basic properties of zero modified Poisson (ZMP) and zero modified negative binomial (ZMNB) processes are described. This section also introduces some Markov sequences suitable for generating the intensities in our ZMSCD models. In Section 3 we formulate the model in GSS form and then write down the GKF algorithm. Section 4 discusses the details of EF method for parameter estimation in our model. Simulation results to illustrate the computation methods are summarized in Section 5. Real  data sets are also   analyzed in Section 6 and some concluding remarks are provided in Section 7.

  \section{ZMSCD models generated by Markovian intensities}
  Let $\{Y_t, t=0, \pm1, \pm2, ...\}$ be a discrete time count process on the state space 
   $ \{0,1,2,...\}$ and  $\cF_{t-1}$ be the sigma field generated by $\{Y_{t-1}, Y_{t-2},...\}$. 
     Suppose that, conditional on $\cF_{t-1},$ the rv $Y_t$ follows a zero-modified distribution (ZMD) with stochastic intensity function $\la_t$. We assume that the stochastic intensity, $\{\la_t\}$ is a stationary Markov sequence of non-negative rvs. Let us now describe two forms of ZMD which were introduced to study the count regression models. The first one initially studied by Dietz and Bohning (2000) for a zero modified Poisson model and then generalized by Bertoli et al (2019) is defined by  
 \benr \lel{zmscd1}
 P^{M}(Y_t = k|\cF_{t-1},p_t,\la_t) =
 \begin{cases}
 	(1 - p_t) + p_tP(Y_t = 0 |\cF_{t-1},\la _t), & k = 0 \\
 	p_tP(Y_t = k|\cF_{t - 1},\la _t), & k > 0, \quad \quad \quad
 \end{cases} 
 \eenr
 where $p_t$ is the  zero modification parameter such that
 \benrr 
 0 \le {p_t} \le P^{-1}(Y_t = 0|\cF_{t-1},\la _t).
 \eenrr
  The model is quite general in the sense it includes several special cases. For example, if
  \begin{enumerate}
  	\item $p_t=0$, then $P^M(Y_t=0|p_t,\lambda_t)=1$, which gives the PMF of a rv degenerate at $0$.
  	\item $p_t=1$, then $P^M(Y_t=0|p_t,\lambda_t)=P(Y_t=0|\lambda_t)$ which is the usual model without any zero modification.
  	\item $0\le p_t \le 1$, then $(1-p_t)P(Y_t=0|\lambda_t)>0,$ which implies that the modified distribution has more zeros than the zeros in the base line distribution and hence this is a case of zero  inflation.
  	\item ${p_t} \in \left[ {1,{P^{ - 1}}\left( {\left. {{Y_t} = 0} \right|{\cF_{t-1}},{\lambda _t}} \right)} \right]$, then $(1-p_t)P(Y_t=0|\lambda_t)<0$ which is a case of zero deflation.
  	\item ${p_t} = {P^{ - 1}}\left( {\left. {{Y_t} = 0} \right|{\cF_{t-1}},{\lambda _t}} \right)$, then $P(Y_t=0|\lambda_t)=0$ which implies the zero truncated case with pmf:
  \benrr	
  \hspace{-0.2in}	P_T( Y_t = k|\cF_{t-1},p_t,\la_t ) 
  	&=& \frac{P(Y_t = k |{\cF_{t-1}},\la_t)}{P(Y_t > 0 |{\cF_{t-1},\la_t } )}( {1 - {\del _Y}}),\ \
  	\text{ where} \ \ \
  \del _Y = 
  \begin{cases}
  	1\quad & if\quad {Y_t} = 0\\
  	0\quad & \text{otherwise}.
  \end{cases}
 \eenrr
  \end{enumerate}
In short, one can easily see that (\ref{zmscd1}) is not a kind of mixture distribution typically chosen to handle the zero inflated data. Also, it is straight forward to see that $P^M(Y_t|p_t,\lambda_t) \ge 0$ and is a proper PMF. To see the effect of zero modification parameter $p_t$, the proportion of additional or missing  zeros can be computed as  \[{P^M}\left( {\left. {{Y_t} = 0} \right|{\cF_{t-1}},{p_t},{\lambda _t}} \right) - P\left( {\left. {{Y_t} = 0} \right|{\cF_{t-1}},{\lambda _t}} \right) = \left( {1 - {p_t}} \right)P\left( {\left. {{Y_t} > 0} \right|{\cF_{t-1}},{\lambda _t}} \right).\]
So for a specified value of $p_t$, one can identify the nature of zero modifications with respect to the base distribution.

An alternative parameterization of the ZMD proposed by Barreto-Souza (2015) is, 
\benr \lel{zmscd2}
P^{M}(Y_t = k|\cF_{t-1},p_t,\la_t) =
\begin{cases}
	p_t +(1- p_t)P(Y_t = 0 |\cF_{t-1},\la _t), & k = 0 \\
	(1-p_t)P(Y_t = k|\cF_{t-1},\la _t), & k > 0, \quad \quad \quad
\end{cases} 
\eenr
In this parameterziation, the range of zero modification parameter $p_t$ is specified by 
\benrr
 - \frac{{P\left( {\left. {{Y_t} = 0} \right|{\cF_{t-1}},{\lambda _t}} \right)}}{{1 - P\left( {\left. {{Y_t} = 0} \right|{\cF_{t-1}},{\lambda _t}} \right)}} \le {p_t} \le 1.
\eenrr
%
This representation also covers some of the interesting special cases. For example, if 
\begin{enumerate}
	\item $p_t=0$, then $P^M(Y_t=0|p_t,\lambda_t)=P(Y_t=0|\lambda_t)$ which gives usual model witout any zero modification.
	\item $p_t=1$, then $P^M(Y_t=0|p_t,\lambda_t)=1$, which is the case of degeneracy at 0.
	\item $0\le p_t \le 1$, then $p_tP(Y_t=0|\lambda_t)>0$ implies the modified distribution has an excess proportion of zeros larger than the base line distribution and hence there is zero  inflation.
	\item ${p_t} \in \left( { - \frac{{P\left( {\left. {{Y_t} = 0} \right|{\cF_{t-1}},{\lambda _t}} \right)}}{{1 - P\left( {\left. {{Y_t} = 0} \right|{\cF_{t-1}},{\lambda _t}} \right)}},0} \right)$, then $p_tP(Y_t=0|\lambda_t)<0,$ which implies that the modified distribution has lesser proportion of zeros  than the base line distribution and hence it is a case of zero  deflation.
	\item ${p_t} =  - \frac{{P\left( {\left. {{Y_t} = 0} \right|{\cF_{t-1}},{\lambda _t}} \right)}}{{1 - P\left( {\left. {{Y_t} = 0} \right|{\cF_{t-1}},{\lambda _t}} \right)}}$, then $P(Y_t=0|\lambda_t)=0$ which implies the zero truncated case.  
\end{enumerate}

Though both these forms (\r{zmscd1} and \r{zmscd2}) of ZMD are used in the literature, we consider the second form (\r{zmscd2}) in this paper. 
Our objective here is to study the properties of ZMSCD models when their intensity functions are generated by stationary non-negative Markov sequences. Further we assume that the zero modification parameter $p_t=p$ as constant.   
  By the basic property of parameter driven models for counting processes, $Cov(Y_t|\cF_{t},Y_s|\cF_{s}) =0$ for $s\ne t$.
  The following notations are introduced for a stationary Markov sequence $\{\la_t\}$. Let $\mu_{\la} = E(\la_t), \si_{\la}^2= Var(\la_t)$ and the $lag$-$k$ autocovariance and autocorrelation (ACF) of $\{\la_t\}$  are respectively denoted by $ \g_{\la}(k) = Cov(\la_t,\la_{t+k})$  and $\rho_{\la}(k) =\g_{\la}(k)/\si_{\la}^2 .$
 We focus on two of the commonly used zero modified counting processes namely zero modified Poisson (ZMP) and zero modified negative binomial (ZMNB). Other zero modified models  are also possible  based on Generalized Poisson distribution, Poisson Lindley distribution, Poisson Inverse Gaussian distribution, geometric distribution, etc. Some of the elementary properties of the resulting ZMSCD models are described below. Zhu(2012) studied the details of zero inflated observation driven models.
 
 \subsection{Zero Modified Poisson SCD Model}
 We say that $\{Y_t\}$ is a zero modified Poison stochastic conditional duration (ZMPSCD) process if  the conditional distribution of $Y_t$ given the past information follows a zero modified Poisson distribution defined by  
  \benr \lel{zip}
P(Y_t=k|\cF_{t-1}) =
\begin{cases}  
\omega_t + (1-\omega_t) e^{-\la_t} \ \ \ \  \ \text{if} \ \ \ \ \ \ k=0,	 \\
(1-\omega_t)e^{-\la_t} \la_t^k /k! \ \ \ \  \text{if} \ \ \ \ \ \ k=1,2,...,
\end{cases}
 \eenr
where $\{\la_t\}$ is a stationary Markov sequence of non-negative rvs.
This model is obtained from the ZMSCD model (\r{zmscd2}) by taking $p_t=\omega_t, 0 < 1-{p _t} < \frac{{{e^{{\lambda _t}}}}}{{{e^{{\lambda _t}}} - 1}}$ 
and the base line distribution as the Poisson with mean $\lambda_t$. When $\omega_t=\omega$ and the range of $\omega$ is restricted to the unit interval, the model reduces to Zero Inflated Poisson(ZIP) model.  The conditional moments of any order $Y_t$ given $\cF_{t-1}$ can be obtained using its probabilty generating function (PGF): $G_y(s)=\omega+(1-\omega)e^{(s-1)\la_t}$, for $0<s<1$. Thus we have the conditional mean and variance respectively: 
\benrr
E(Y_t|\cF_{t-1}) = (1-\omega)\la_t \ {\text{and}}\ \ \Var(Y_t|\cF_{t-1}) = (1-\omega)(1+\omega\la_t)\la_t.
\eenrr
Consequently the unconditional mean and variance of $Y_t$ respectively become
\benrr
E(Y_t) = E(E(Y_t|\cF_{t-1})) = E((1-\omega)\la_t) = (1-\omega) \mu_{\la}\\
V(Y_t) =
 V(E(Y_t|\cF_{t-1}))+E(V(Y_t|\cF_{t-1})) 
= (1-\omega)[\mu_{\la}+\si_{\la}^2 + \omega \mu_{\la}^2].
\eenrr  
To compute autocovarance function of $\{Y_t\}$ consider
\benrr
E(Y_tY_{t+k}) 
=E(E(Y_tY_{t+k}|\cF_{t-1},\cF_{t+k-1} ))
= (1-\omega)^2 E(\la_t.\la_{t+k}),
\eenrr
where we have used the conditional independence of $Y_t$ given the past.
 Thus, we have the autocovarance function of $\{Y_t\}$:
\benrr
\g_y(k) = Cov(Y_t, Y_{t+k}) = (1-\omega)^2 \g_{\la}(k)
\eenrr
and its autocorrelation function (ACF) : 
\benrr
\ro_y(k) = \frac{(1-\omega)^2 \g_{\la}(k)}{(1-\omega)[\mu_{\la}+\si_{\la}^2 + \omega \mu_{\la}^2]}  = \frac{(1-\omega)\si_{\la}^2\ro_{\la}(k)}{\mu_{\la}+\si_{\la}^2 + \omega \mu_{\la}^2} \le  \ro_{\la}(k).
\eenrr
The geometrically decreasing behaviour of the ACF of a stationary Markov sequence $\{\la_t\}$ is preserved by the corresponding ZMSCD sequence. However, $\ro_y(k) \le  \ro_{\la}(k)$ for every $k$ over the whole parameter space.  Fig.1a is a sample plot to show this behaviour.

  \subsection{Zero modified negative binomial SCD Models}
  In addition to zero modification, overdispersion can also be present in many count time series. The zero modified negative binomial (ZMNB) model can be used in such situations. 
   The ZMNBSCD model is obtained from the general ZMSCD model, (\r{zmscd2}) by taking $p_t=\omega_t$ such that  \[0 < 1-{p_t} < {\left[ {1 - {{\left( {\frac{1}{{1 + a\lambda _t^c}}} \right)}^{\frac{{\lambda _t^{1 - c}}}{a}}}} \right]^{ - 1}}\] and the base line distribution as the negative binomial with parameters  $\lambda_t>0$ and $a>0$.

  In terms of notations of Zhu(2012), the probability mass function of ZMNB distribution with intensity function $\la_t$ is given by 
    \benr \lel{zinb}
  P(Y_t=k|\cF_{t-1}) =
  \begin{cases}
  	\omega_t+(1-\omega_t)\Big(\frac{1}{1+a\la_t^c}\Big)^{\la_t^{1-c}/a}, & \text{if}\ \  k=0,\\
  	 (1-\omega_t)\frac{\G(k+\la_t^{1-c}/a)}{k!\G(\la_t^{1-c}/a)} \Big(\frac{1}{1+a\la_t^c}\Big)^{\la_t^{1-c}/a} \Big(\frac{a\la_t^c}{1+a\la_t^c}\Big)^k,&  \text{if} \ \ k=1,2,...  ,	
  \end{cases}
  \eenr
  where $\la_t>0,  a > 0$ and we denote it by $(Y_t=k|\cF_{t-1}) \sim ZMNB(\la_t, a, \omega_t).$ The index $c=0,1$ identifies the particular form of the negative binomial distribution. When $\omega_t=\omega$ and $0<\omega<1$, the model becomes zero inflated negative binomial model. The conditional and unconditional moments of such model are similar to those in the case of ZMPSCD model, which are listed below:
  \benrr
  E(Y_t|\cF_{t-1}) = (1-\omega)\la_t \ \ \Var(Y_t|\cF_{t-1}) = (1-\omega)(1+\omega\la_t+a\la_t^c)\la_t, \ \  E(Y_t) = (1-\omega)\mu_{\la},
  \eenrr
  \benrr
  V(Y_t) =
  \begin{cases}
   (1-\omega)[(1+a)\mu_{\la}+\si_{\la}^2 + \omega \mu_{\la}^2]  \ \ \ \ \text{if} \ \ \ \ \ \ c=0,\\
    (1-\omega)[\mu_{\la}+(a+1)\si_{\la}^2 + (\omega +a)\mu_{\la}^2] \ \ \ \ \text{if} \ \ \ \ \ \ c=1. 
  \end{cases}
  \eenrr 
  Further the ACF may be expressed as 
 \benrr
 \rho_y(k) =
 \begin{cases}
 	\frac{(1-\omega)\si_{\la}^2\rho_{\la}(k)}{[(1+a)\mu_{\la}+\si_{\la}^2 + \omega \mu_{\la}^2] } \ \ & \text{if} \ \ \ \ \ \ c=0,\\
 	\frac{(1-\omega)\si_{\la}^2\rho_{\la}(k)}{[\mu_{\la}+(a+1)\si_{\la}^2 + (\omega +a)\mu_{\la}^2]} \ \ \ \ & \text{if} \ \ \ \ \ \ c=1. 
 \end{cases}
 \eenrr 
  	\begin{center} 
  		\framebox{  \textbf{Figure 1  about here}.}
  	\end{center}
%
%
%
%
 {\bf{Note 1:}} As $a\to 0$, the $ZMNB(\la_t, a, \omega)$ reduces to  $ZMP(\la_t, \omega)$. Consequently, the mean, variance and ACF of ZMNB count process reduce to those of ZMP count process as can be easily verified from the corresponding expressions.\\
 The following fourth moment for the ZMNB process is useful for estimating the parameters. 
 \benr \lel{ZINBFourth}
 E(Y_t -(1-\omega)\la_t)^4|\cF_{t-1})
 &=& (1-\omega)\la_t\Big[\la_t^3(3\omega^3-3\omega^2+\omega) + \nn 6\la_t^2\omega^2+4\la_t\omega+3\la_t+1 + 6a^3\la_t^{3c}\\
 && + (12a^2+(3+8\omega)a^2\la_t)\la_t^{2c} + (6a\la_t^2\omega^2+6(1+2\omega)a\la_t)\la_t^{c}\Big].
 \eenr
  \subsection {Examples for Markov sequences of intensities.} \lel{sec2.3}
 {\bf{ 1. First order autoregressive (AR(1)) model for non-negative rvs:}} Let $\{\eta_t\}$ be a sequence of iid rvs with $E(\eta_t) = \mu_{\eta}$ and $V(\eta_t) = \si^2_{\eta}$. Define
  \benr \lel{ar1}
  \la_t =\rho \la_{t-1} +\eta_t,      \ \ \ \  0\le \rho<1.
  \eenr
  The distribution of $\eta_t$ is chosen in such a way that  $\{\la_t\}$ defines a stationary sequence of specified marginal distribution.  
  In particular this model includes the exponential AR(1) (EAR(1)) model and gamma AR(1) (GAR(1)) model defined by Gaver and Lewis (1980). In the case of EAR(1) model, the marginal distribution of $\{\la_t\}$ is exponential with pdf:
  \benr\lel{Exp}
  f(x;\b) = \b e^{-\b x}, x\ge 0, \b >0
  \eenr
 if and only if the distribution function of the innovation rv, $\eta_t$ in (\r{ar1}) is given by 
 \benr\lel{EAR1-inno}
 F_{\eta}(x) = \rho + (1-\rho)(1-e^{-\b x}), x\ge 0, \b >0.
 \eenr
 The intensity sequence, $\{\la_t\}$ is a GAR(1) sequence if each $\la_t$ follows a gamma marginal ($G(\b,p)$) distribution with pdf: 
  \benr\lel{gamma}
  f(x|\b,p) = \frac{e^{-\b x}\b^p x^{p-1}}{\G(p)},  \ \ \ x\ge 0, \b>0, p>0.
  \eenr
 The marginal distribution of $\{\la_t\}$ in (\r{ar1}) is $G(\b,p)$  if and only if the distribution of the innovation rv, $\eta_t$ is given by (cf, Lawrance (1982)) that of 
$ \eta_t = \sum_{i=1}^{N}\rho^{U_i}E_i,$ 
where $\{U_i\}$ and  $\{E_i\}$ are mutually independent iid $Unif(0,1)$ and $Exp(\b)$ rvs  and $N$ follows a Poisson distribution with mean $plog(1/\rho).$ \\

{\bf{2. Random Coefficient AR(1) (RCAR(1)) models:}} Let $\{J_i\}$ be a sequence of iid rvs distributed over $(0,1)$ and difine 
 \benr \lel{rcar1}
\la_t =J_{t} \la_{t-1} +\eta_t,      \ \ \ \  0\le E(J_t)<1.
\eenr
The gamma Markov sequences defined by Sim (1990) and Beta Gamma AR(1) model difined by Lewis et al (1989) are included in the above RCAR(1) model. The NEAR(1) and TEAR(1) models of Lawrance and Lewis (1981) are also special cases of (\r{rcar1}).  

{\bf{3. Product AR(1) models:}} Let $\{V_t\}$ be a sequence of iid non-negative rvs with $E(V_t) = \mu_{v}$ and $V(V_t) = \si^2_{v}$. Assume that $\la_0$ is independent of $V_1$ and define 
 \benr \lel{par1}
\la_t = \la^{\rho}_{t-1} V_t,      \ \ \ \  0\le \rho <1, t=1,2,....
\eenr
 Mckenzie (1982) introduced this model for defining a stationary gamma Markov sequence and compared it with the GAR(1) model. 
Abraham and Balakrishna (2012) obtained an explicit form of the distribution of the innovation rv, $V_t$.  
Muhammed, Balakrishna and Abraham (2019) obtained the innovation distribution of generalized gamma PAR(1) model and proposed methods of estimation for the stochastc volatility models generated by it. 
%

{\bf{4. Pitt-Walker models:}} Pitt and Walker (2005) proposed a method of constructing stationary AR(1) sequences $\{\la_t\},$ by choosing marginal and conditional distributions which satisfy a particular equation. They used this idea to construct stationary Markov sequences with stationary marginal distributions such as gamma, inverse gamma, etc. The resulting sequence satisfies the relation: $E(\la_t|\la_{t-1}) = \rho\la_{t-1} + (1-\rho)\mu_{\la},\ \ \ 0\le\rho<1.$

\section{State Space representation of ZMSCD models }
As stated in Section 1, we propose EF method for parameter estimation, which requires the filtering of the latent intensities. This can be facilitated by expressing the model in a generalized state space (GSS) form and then using  generalized Kalman filtering (GKF). 
  To achieve this goal we adapt the method proposed by Zehnwirth (1988), which is summarised below for our reference. The GSS model contains an observation equation in terms of the data and a state equation.
 \benr \lel{OE}
  Y_t=F_t\Lambda_t + \vep_t \ \ \ \  \text{: observation equation} 
 \eenr
 \vspace{-1cm}
 \benr \lel{SE}
 \Lambda_t = G_t \Lambda_{t-1} + W_t, \ \ \  \text{ : state equation,}
 \eenr
  where (a) $\Lambda_t$ is a q-dimensional state vector; (b) $F_t$, and $G_t$
 are known matrices of dimensions $p \times q$ and $q \times q$, respectively;
 (c) $\vep_t$, is a p-dimensional observation error such
 that $E[\vep_t|\Lambda_t] = 0$, $Cov[\vep_t|\Lambda_t] = V_t(\Lambda_t),$ where $V_t(.)$ is a  known function of unknown $\Lambda_t$ and
 $Cov[\vep_t,\vep_s|\Lambda_t,\Lambda_s] = 0$ for all $s \ne t$. 
  (d) The q-dimensional random vectors $W_t$, form an uncorrelated
 sequence with $E[W_t] = 0$ and $Cov[W_t] = w^*_t$ with ${w^*_t}$ known, and $C[W_t,\Lambda_s] = 0$ for $t > s$; and (e) $Cov(\vep_t,W_s|\Lambda_t) = 0$ for
 all $s$ and $t$. Consequently it follows that $Cov[Y_t,Y_s|\Lambda_t] = 0$
 for all $t > s.$ Under this set up Zehnwirth (1988), established the following GKF algorithm for the GSS model, where $\h\Lambda_t$ denotes the filtered value of $\Lambda_t$ conditional on the past observations:
 \benr \lel{GKF}
 \hat\Lambda_{t|t-1} = G_t\hat\Lambda_{t-1}, \\
 \hat\Lambda_{t} = \hat\Lambda_{t|t-1} + K_t(Y_t - F_t\hat\Lambda_{t|t-1}), \\
 K_t = C_{t|t-1}F_t'[F_tC_{t|t-1}F_t' + \bar V_t]^{-1}, \\
 C_t = [I - K_tF_t]C_{t|t-1},
 \eenr
where $\hat\Lambda_{t|s}$ is the minimum mean squared error linear estimator
of $\hat\Lambda_t$ based on $Y_s, (s\le t)$, $C_{t|s}$ is the unconditional
error covariance matrix of $\hat\Lambda_{t|s}$ and $\bar V_t = E(V_t(\Lambda_t))$. In particular, one writes $\hat\Lambda_{t} = \hat\Lambda_{t|t}$ and $C_{t} =C_{t|t}.$
The major difference between GSS and the usual SS set up is that $V_t(.)$ is a known constant in the latter case. But in GSS model, $V_t(\Lambda_t)$ is a known function of unknown $\Lambda_t$. In order to implement the above GKF we replace $V_t(\Lambda_t)$ by its expectation $\bar V_t$, computed using the state equation.  Next we simplify the algorithm when $\{Y_t\}$ follows a zero-modified SCD model with intensities  generated by certain non-negative Markov sequences
 described in Section 2, where $\la_t$ denotes the one-dimensional intensity function.
 The filtering algorithms also lead to suitable estimating equations to estimate the parameters inolved, which we explore in the next section. See also Thavaneswaran et al. (2015) and Thekke et al. (2016). 
\section{Estimation for Zero modified SCD models}  

As discussed in Section 2, conditional on the past, $Y_t$ follows a zero modified distribution with intensity function $\la_t$. 
Further we assume that $\{\la_t\}$ follows a stationary Markov sequence.  Let $\boldsymbol{\th} =(\th_1,\th_2,...,\th_p)',$ (where prime denotes the transpose of a vector) be the vector of parameters indexing the finite dimensional distribution of $\{\la_t\}$. There are no parameters other than $\boldsymbol{\th}$ and $\omega$ in the whole model. So the components of $\boldsymbol{\th}$ are the parameters present in the stationary marginal distribution of $\{\la_t\}$ and its one-step transition distribution. Based on the theory of linear estimating equations for stochastic processes, one may estimate the parameters of interest using the optimal EF based on the following martingale EFs: 
 \benr \lel{ef12}
 g_{1t} = Y_t - E(Y_t|\cF_{t-1}) = Y_t - (1-\omega)\la_t \\
 g_{2t} = \la_t - E(\la_t|\cF_{t-1}) = \la_t - E(\la_t |\la_{t-1}).
 \eenr 
 Note that in the models listed above, $E(\la_t|\cF_{t-1})$ is a  (linear or nonlinear) function of $\la_{t-1}$ and the parameters. 
 After establishing the optimal properties of the estimating functions, we can find the estimates by solving the optimal estimating equations based on  $g_{1t}, g_{2t}$. But, $g_{it}, i=1,2 $ contain the latent variables $\la_t$, which are not observable. So we implement the methods discussed in Section 3 to  filter these latent variables and use them for estimation. In fact the GKF alogorithm leads to a useful estimating equation, which may be used for estimation. 
 We illustrate the proposed methods under two special cases, namely ZMPSCD and ZMNBSCD which are discussed in the following subsections.
 
 \subsection{Estimation for ZMPSCD model}
 Under a ZMPSCD model, we have
 $E(Y_t|\cF_{t-1}) = (1-\omega)\la_t$ and $V(Y_t|\cF_{t-1}) = (1-\omega)(1+\omega\la_t)\la_t$. 
 Accordingly the GKF algorithm described in (3.3) to (3.6), may be written as: 
 \benr \lel{gkf}
 \hat\la_{t} =  \hat\la_{t|t-1} + \frac{(1-\omega)C_{t|t-1}}{(1-\omega)^2C_{t|t-1} +(1-\omega) v}(Y_t - (1-\omega)\hat\la_{t|t-1})
 \eenr
  with   $v=\mu_{\la}+\omega(\si_{\la}^2+\mu_{\la}^2).$
  That is,
  \benr\lel{ZIPfilter}
  \lefteqn{
  \hat\la_{t|t} = \rho \hat\la_{t-1|t-1} + (1-\rho)\mu_{\la}} \nn \\
  && \hspace{.2in}+ \frac{(1-\omega)[\rho^2C_{t-1|t-1}+(1-\rho^2)\si_{\la}^2]}{(1-\omega)^2[\rho^2C_{t-1|t-1}+(1-\rho^2)\si_{\la}^2] +(1-\omega) v}(Y_t - (1-\omega)[\rho \hat\la_{t-1|t-1} + (1-\rho)\mu_{\la}]).
  \eenr
  This equation helps in getting filtered values of $\la_t$ provided the parameters and the initial values are known. The suggested values for the initial values above are also in terms of parameters. Next we will identify suitable optimal estimating equations in terms of the filtered values to estimate the unknown parameters. Intoducing notations:
  \benrr
   L_t=\hat\la_{t|t},
   h_t= Y_t - (1-\omega)[\rho \hat\la_{t-1|t-1} + (1-\rho)\mu_{\la}], 
    P_t=(1-\omega)[\rho^2C_{t-1|t-1}+(1-\rho^2)\si_{\la}^2], \\
    J_t=(1-\omega) (P_t+v), K_t=\frac{P_t}{J_t} \ \  \text{and}\ \  D_t = L_t - \rho L_{t-1} + (1-\rho)\mu_{\la}
  \eenrr
  we can express (\r{ZIPfilter}) as 
$ h_t = D_t\frac{J_t}{P_t}.$
  Clearly 
  \benrr
  E(h_t|\cF_{t-1}) = 0\ \  \text{and}\ \  E(h_t^2|\cF_{t-1}) = V(h_t|\cF_{t-1}) = \frac{J_t^2}{P_t^2}V(D_t|\cF_{t-1}) = \frac{J_t^2}{P_t^2}\frac{P_t}{1-\omega} = \frac{J_t^2}{(1-\omega) P_t}. 
  \eenrr

 To illustrate the method of computation, let us consider the ZMPSCD model in which  $\{\la_t\}$ is a stationary AR(1) sequence of non-negative rvs. In this case the parameter vector to be estimated is 
\benrr 
 \boldsymbol{\th} = (\omega,\mu_{\la},\rho, \si_{\la}^2)'= (\th_1,\th_2,\th_3,\th_4 )' \ \ \text{say}. 
 \eenrr
 We propose to estimate the first three components, namely $\omega,\mu_{\la},\rho$ using an optimum unbiased estimating function based on a martingale sequence:
 \benrr
 h_t= Y_t - (1-\omega)[\rho \hat\la_{t-1|t-1} + (1-\rho)\mu_{\la}],
 \eenrr
  defined above. The fourth component, $\si_{\la}^2$ will be estimated based on the resulting residuals.  To begin with let $\cG = \Big\{g=(g^{(1)}, g^{(2)}, g^{(3)})'= \st a_{t-1}h_t\Big\}$	be the class of linear unbiased estimating functions. Then from the theory of martingale estimating functions (cf, Godambe (1985)), the optimal EF is given by 
 $ g^*=(g^{*(1)}, g^{*(2)}, g^{*(3)})'$ with 
 \benrr
 g^{*(i)} = \st a^{*(i)}_{t-1}h_t \ \ \text{and} \ \
  a^{*(i)}_{t-1} = \frac{E\Big(\frac{\partial h_t}{\partial \th_i}|\cF_{t-1}\Big)}{Var(h_t|\cF_{t-1})}\ \ \text{for}\ \ i=1,2,3.
  \eenrr
For the above martingale sequence $\{h_t\}$ we have, \\
  $ \frac{\partial h_t}{\partial \th_1} = \rho \hat\la_{t-1} + (1-\rho)\mu_{\la},\ \ 
    \frac{\partial h_t}{\partial \th_2} = -(1-\omega)(1-\rho),\ \ 
  \frac{\partial h_t}{\partial \th_3} = -(1-\omega)( \hat\la_{t-1} -\mu_{\la}).$\\
  Since $ \hat\la_{t-1}$ is measurable with respect to $\cF_{t-1}$, it follows that  
  \benrr
   E\Big(\frac{\partial h_t}{\partial \th_i}|\cF_{t-1}\Big) = \frac{\partial h_t}{\partial \th_i} \ \  \text{and} \ \  
  a^{*(i)}_{t-1} = \frac{(1-\omega)P_t}{J_t^2}\frac{\partial h_t}{\partial \th_i}, i= 1,2,3. 
  \eenrr
  Hence the optimal EF is given by $ g^*=(g^{*(1)}, g^{*(2)}, g^{*(3)})'$ where 
  \benrr
  g^{*(1)} = \st \frac{(1-\omega)P_t}{J_t^2}[\rho \hat\la_{t-1} + (1-\rho)\mu_{\la}](Y_t - (1-\omega)[\rho \hat\la_{t-1} + (1-\rho)\mu_{\la}])\\
  g^{*(2)} = \st \frac{(1-\omega)P_t}{J_t^2}[-(1-\omega)(1-\rho)](Y_t - (1-\omega)[\rho \hat\la_{t-1} + (1-\rho)\mu_{\la}])\\
   g^{*(3)} = \st \frac{(1-\omega)P_t}{J_t^2}[-(1-\omega)( \hat\la_{t-1} -\mu_{\la})](Y_t - (1-\omega)[\rho \hat\la_{t-1} + (1-\rho)\mu_{\la}]).  
  \eenrr
 The estimates are obtained by solving the equation $g^*=\bf{0},$ which requires a suitable iterative method.  We suggest ordinary moment estimates for the initial values to implement the iterative procedure. 
 In order to estimate $\si_{\la}^2$, we propose the EF based on 
 \benr \lel{h2t}
 h_{2t} =\Big[\big(\la_t - \rho \la_{t-1} - (1-\rho)\mu_{\la}\big)^2/(1-\rho^2) -\si_{\la}^2 \Big].
 \eenr
 Clearly, $E(h_{2t}|\cF_{t-1}) = 0$ and $Var(h_{2t}|\cF_{t-1}) =E\Big(\la_t - \rho \la_{t-1} - (1-\rho)\mu_{\la}\Big)^4/((1-\rho^2))^2 -\si_{\la}^4.$ Now the optimal estimating function is chosen from the class $\cG_1 = \Big\{g= \st b_{t-1}h_{2t}\Big\}$ with optimum coefficients
 $b^*_{t-1}=\frac{E\Big(\frac{\partial h_{2t}}{\partial \th_4}|\cF_{t-1}\Big)}{Var(h_{2t}|\cF_{t-1})} = \frac{-1}{Var(h_{2t}|\cF_{t-1})}.$
  For evaluating the estimate, we can replace $\la_t $ by its filtered value and $\rho$ and $\mu_{\la}$ by the respective estimates. If $\{\la_t\}$ is a stationary homoscedastic Markov sequence like the $GAR(1)$ then  $b^*_{t-1}$ will be a constant  
%
and hence the estimate of $\si^2_{\la}$ will be
	\benr \lel{varest}
	 \hat\si^2_{\la} = \frac{1}{n} \st\Big(\la_t - \rho \la_{t-1} - (1-\rho)\mu_{\la}\Big)^2/(1-\rho^2). 
\eenr

 In particular if $\{\la_t\}$ a GAR(1) process, its acf is  $\ro_{\la}(k)=\ro^k$. Let $\bar Y$, $s^2$ and $r_1$ be the sample mean, sample variance and the first order sample acf of $\{Y_t\}$.  Equating them to the corresponding moments of $\{Y_t\}$ in Section 2.1, we can write 
 \[
 \begin{array}{l}
 \bar Y = (1 - \omega)\mu _\lambda  ,\quad s^2  = (1 - \omega)\left[ {\mu _\lambda   + \sigma _\lambda ^2  +\omega \mu_{\la} ^2 } \right], \quad 
 r_1  = \frac{{(1 - \omega)\sigma _\lambda ^2 \rho }}{{\mu _\lambda   + \sigma _\lambda ^2  +\omega \mu_{\la} ^2 }}. \\ 
 \end{array}
 \]
 Simultaneous solution of these equations will provide a set of moment estimates of the model parameters to initialize the computations. The details are dscribed in the following algorithm. 
 If $p=1$, we get the intensities generated by an Exponential AR(1) (EAR(1)) model and the corresponding expressions reduce to the following.  
 \benrr
 \bar Y = (1-\omega)\mu_{\la}, s^2 = (1-\omega)\mu_{\la}[1 + (1+\omega) \mu_{\la}], 
 r_1=\frac{(1-\omega)\mu_{\la}\ro}{1+\mu_{\la}(1 + \omega)}.
 \eenrr
 Superfixing (0) to denote the initial values and then simplifying, we get 
 \benrr
 \omega^{(0)} = \th^{(0)}_1 = \frac{z-1}{z+1},  {\text{with}}\ \  z=\frac{1}{\bar Y}(\frac{s^2}{\bar Y}-1)\\ 
 \mu^{(0)}_{\la} =  \th^{(0)}_2 = \frac{\bar Y}{(1-\omega)},\ \ \
 \ro^{(0)} = \th^{(0)}_3 = r_1\frac{1 + (1+\omega) \mu_{\la}}{(1-\omega) \mu_{\la}}.
 \eenrr

The following algorithm may be used for filtering the stochastic intensity and estimate the parameters iteratively for a ZMP model with GAR(1) intensities.\\
 {\bf{Filtering and estimation algorithm:}} 
 \begin{enumerate}
 	\item Initialize $\th^{(0)}_i, i=1,2,3,$ and $C_{1|0}= (1-\rho^2)\si_{\la}^2 $
 	and $\hat\la_{1|0} = \rho \la_0 + (1-\rho)\mu_{\la}, {\la_0\sim G(\beta, p)} $. 
 	\item Use (\r{gkf}) and (\r{ZIPfilter}) for updating $\la$- values
 	\benr \lel{Pfilter}
 	\hat\la_{t} =  \hat\la_{t|t-1} + \frac{(1-\omega)C_{t|t-1}}{(1-\omega)^2C_{t|t-1} +(1-\omega) v}(Y_t - (1-\omega)\hat\la_{t|t-1}).
 	\eenr
 	\item Compute $\th^{(k)}_i, i=1,2,3; k=1,2,...$ by solving the estimating equations $g^{*}=\bf{0}$.
 	\item Store the filtered values of $\la_t$ to obtain $\hat\si^2_{\la}$ using (\r{varest}) when $\la_t$ are generated by the GAR(1) model.
 	\item The parameters $\b$ and $p$ may be estimated by $\hat\b = \hat\mu_{\la} /\hat\si^2_{\la} $ and $\hat p=\hat\mu_{\la} \hat\b$.
 	\item  Repeat the above steps with $k=k+1$ until convergence. 
 
 	\end{enumerate}
 
 \subsection{Estimation for ZMNBSCD model}
 
 Suppose that $Y_t$ given the past, follows a ZMNB distribution. From Section 2.2, we have 
 \benrr
  E(Y_t|\cF_{t-1}) = (1-\omega)\la_t \ \  {\text{and}} \ \  \Var(Y_t|\cF_{t-1}) = (1-\omega)(1+\omega\la_t + a\la_t^c)\la_t. 
 \eenrr  
 If the sequence of intensities $\{\la_t\}$ is generated by a stationary non-negative AR(1) model then the GKF system described in Section 3 will remain same except for the expression of $\bar V_t,$ which is given by $V_t(\lambda_t)$.  
 In terms of notations of Section 3, we have
 \benr\lel{vtbarNB}
 \bar V_t 
 &=& E(Var(Y_t|\cF_{t-1})) \nn \\
 &=& 
 \begin{cases}
  (1-\omega)[(1+a)\mu_{\la} + \omega(\si_{\la}^2+\mu_{\la}^2)], & \ \ \  {\text{if}} \ \  c=0   \\
  (1-\omega)[\mu_{\la} + (\omega+a)(\si_{\la}^2+\mu_{\la}^2)], & \ \ \ {\text{if}} \ \ c=1.
 \end{cases} =  (1-\omega)v_b^{(c)}, {\text{say}}.
\eenr
This $\bar V_t$ reduces to that of ZIPSCD model if $a=0$. 

The value of $c$ is taken as either 0 or 1 according to the selected class of ZMNB distribution. So the equation for filtering $\la_t$ is same as (\r{ZIPfilter}) with $v$ replaced by $v_b^{(c)}$ given in (\r{vtbarNB}) . The parameter vector to be estimated here is $\boldsymbol{\th}= (\omega,\mu_{\la},\rho,a, \si_{\la}^2)'= (\th_1,\th_2,\th_3,\th_4, \th_5)'$ say. The first three components, namely $\omega,\mu_{\la},\rho$ can be estimated using the EF resulting from equation (\r{ZIPfilter}) by repeating the method described in Section 4.1. For estimating the dispersion parameter $\th_4 = a$, we use the following quadratic EF:
\benrr
h_t^Q 
&=& (Y_t-E(Y_t|\cF_{t-1}))^2 - V(Y_t|\cF_{t-1})\\
&=& (Y_t-(1-\omega)\la_t)^2 - (1-\omega)(1+\omega\la_t + a\la_t^c)\la_t.
\eenrr
Clearly $E(h_t^Q|\cF_{t-1}) = 0$ and
\benr 
Var(h_t^Q|\cF_{t-1}) \lel{varQ}
&=& E[(Y_t-(1-\omega)\la_t)^4|\cF_{t-1}] - (1-\omega)^2(1+\omega\la_t + a\la_t^c)^2\la_t^2 \nn \\ 
&=& (1-\omega)\la_t\Big[\la_t^3(4\omega^3-4\omega^2+\omega) +\la_t^2(8\omega^2-2\omega)+\la_t(5\omega+2)+1+6a^3\la_t^{3c}\\ \nn
&&+\la_t^{2c}(12a^2+2a^2\la_t+9\omega a^2\la_t)+\la_t^{c}(8\omega^2 a\la_t^2 +4a\la_t+14\omega a\la_t - 2\omega a\la_t^2) \Big].
\eenr
Now the optimal estimating function $g^*_Q $ is chosen from the class of EFs  
\benrr
\cG_Q = \Big\{g_Q= \st a^Q_{t-1}h^Q_{t}\Big\} \ \ \text{as} \ \ g^*_Q= \st a^{Q*}_{t-1}h^Q_{t}\\
\text{with}\ \ \ \ \ 
 a^{Q*}_{t-1} = \frac{E\Big(\frac{\partial h^Q_{t}}{\partial a}|\cF_{t-1}\Big)}{Var(h^Q_{t}|\cF_{t-1})} = \frac{-(1-\omega)\la_t\la^c_t}{Var(h^Q_{t}|\cF_{t-1})}.
 \eenrr
 Then obtain the estimate of $a$ as a solution of the equation $g^*_Q= 0.$ Finally for the parameter  $\th_5=\si_{\la}^2$, we propose the same EF based on $h_{2t}$ defined by (\r{h2t}) used for estimating $\si_{\la}^2$ in Section 4.1. The computation algorithm developed in Section 4.1 for ZMPSCD model can be used for ZMNBSCD model by replacing $v$  in (\r{Pfilter}) by $v^{(c)}_b$ given in (\r{vtbarNB}).

\section{Simulation Studies}
A simulation study was conducted to evaluate the finite sample performance of the proposed estimators. First we consider the simulation studies for ZIP model followed by  ZINB model  when intensities are generated by GAR(1) and EAR(1) sequences respectively. That is, we choose the values of zero modification parameter $w$ in the interval $(0,1)$ so that the resulting model becomes suitable to analyze zero inflated count data. The ZMSCD models with intensities generated by other models listed in Section 2.3 may also be considered similarly.  Inspired by the applications of estimating functions in filtering and smoothing of general state space models (cf; Naik-Nimbalkar and Rajarshi, 1995), we have extended their ideas to the case of zero modified time series models. The computational ease of this method facilitates a near optimal solution to a highly non linear and high dimensional numerical integration problem.  The goal of this simulation study is to analyze the sampling behavior of estimators obtained as solution to the appropriate optimal estimating functions rather  than to compare with other methods applicable to ZIP or ZINB models. 
\par In what follows, we simulate a count series of length $n$ from the models, (\r{zip}), (\r{zinb}), (\r{ar1}) for different values of parameters and carry out estimation using the algorithm described in Section 4. 
The moment estimate denoted by $\tilde {\bold \theta}$ is used as initial value to start the algorithm. As there are no closed form expressions for the moment estimates, we applied  a general grid search for initialization.
In some iterations, it is observed that the moment estimates $\tilde w$ and $\tilde \rho$ become infeasible in the sense that they lie outside the parameter space, especially when the experiment is conducted with values of $\rho$ close to 1. In such cases, either we took the initial value from the $\epsilon$ neighborhood of the true value or discard the iteration.  The proportion of infeasible solutions was  relatively small. This iterative process of estimation was repeated 1000 times for each combination of parameters and obtained the estimates and filtered values of latent intensities.  
Table 1 summarizes the simulation results for a ZIPSCD model with GAR(1) intensities.  This Table  gives the average of these 1000 estimates along with the corresponding mean squared error(MSE) in the  parenthesis. Based on our simulation results, we can see that the estimates are close to the true values with small MSEs. The MSE of the estimate of shape parameter $p$ is relatively larger but, when the value of $\rho$ decreases, it also decreases. It is seen that the MSEs decrease when sample size increases.  We observe similar behavior when the intensities are generated by EAR(1) model and hence omit the details.  

\begin{center} 
	\framebox{  \textbf{Table 1  about here}.}
\end{center}
\vspace{0.1in}

\par The details of simulation study for ZINB models when intensities are generated by GAR(1) and EAR(1) sequences are discussed below. 
We consider the case of $c=1$ in the ZINB model given by (\ref{zinb}). In this scenario, the moment estimators do not possess closed form expressions for both EAR(1) and GAR(1) based ZINB models.  Thus, to solve the estimating equations discussed in Section 4.2,  we started  with simple grid search. In Table 2 we summarize the simulation results. 
As in the case of ZIPSCD with GAR(1) intensities, here also the estimates of $\rho$, $\omega$, and $\beta$ performs well in all cases as far as the  bias  and MSEs are concerned. Meanwhile, the estimates of shape parameter $p$ and dispersion parameter $a$, behave little differently, for instance, when $\rho=0.9$, $\omega=0.2$, $\beta=2$, the bias of both $\hat p$ and $\hat a$ become relatively high. This pattern repeats over the other parameter combinations. Though, the MSE of $\hat a$ decreases when $\rho$  increases and vice versa for $\hat p$.
\par The above results show that the performance of    EF based estimators is satisfactory for all combinations of parameters considered  for ZIPSCD with GAR(1) intensities whereas the same is not fully warranted in ZINBSCD models. This observation was also made by Yang et al (2015). They pointed out that, due to the possibility of estimation problems caused by weak identifiability in ZINB models, the dynamic ZINB model is not a good candidate when sample information is limited.   In our case also, the complexity of the ZINBSCD specification may be a reason for  comparatively less efficient estimation results. 
\begin{center} 
	\framebox{  \textbf{Table 2 about here}.}
\end{center}
\vspace{0.1in}

A simulation study is also carried out to see the sampling behavior of the proposed method when a zero deflation is present.  That is, we allow the zero modification parameter $w$ to take negative values so that the resulting model can be used to analyze the zero deflated data.  As in the case of inflated models, we have generated pseudo random numbers of different sizes ($n=200,500, 1000$) from the zero deflated model and applied the estimation and filtering algorithm to obtain the parameter estimates. This procedure is repeated 1000 times and the resulting estimates were saved. Using these values, we have computed the mean and MSE of the estimators.   Table 3 
summarizes the simulation results of ZDPSCD model when a  GAR(1) process is used to generate latent intensities. We skip the results for the sample size $n=500$ to save space.  It is straightforward to see that the bias and MSE decreases as sample size increases.

\begin{center} 
	\framebox{  \textbf{Table 3 about here}.}
\end{center}
\vspace{0.1in}

 Similar pattern was also observed in the case of ZDNBSCD with GAR(1) intensities, presented in Table 4 . However, a weaker performance (in terms of bias and MSE) of the estimates for ZDNBSCD  model is observed compared to that of the ZDPSCD model. This behavior is same as the one  observed for the zero inflated models discussed earlier.

\begin{center} 
	\framebox{  \textbf{Table 4 about here}.}
\end{center}
\vspace{0.1in} 
\section{Data Analysis}
In this Section we analyze two sets of data to illustrate the applications of the proposed models. The first data set consists of weekly number of syphilis cases reported in the state of Maryland in United States from 2007 to 2010. {\it{This data set is  available in the R package ZIM (see Yang et al. 2015)}.} Figure 2 shows the basic structure of the data such as time series, ACF, PACF plots. The histogram clearly indicates that the data contains several zeros. The basic  statistics of the data are given in Table 5. Stationarity of the data is confirmed by performing an augmented Dicky-Fuller test at lag 5, yielding $p$-value 0.01. 
\begin{center} 
	\framebox{  \textbf{Figure 2  about here}.}
\end{center}
\vspace{0.2in}

%
\begin{center} 
	\framebox{  \textbf{Table 5  about here}.}
\end{center}
\vspace{0.2in} 
Observe that the variance is larger than mean implying the presence of over-dispersion.  Also, the sample auto-correlation function and partial auto-correlation functions of the data exhibit temporal  correlations. In fact, the Ljung-Box test for autocorrelation at lag 1 rejects the null hypothesis of zero autocorrelation with $p$-value 0.04104.  To capture all these features,  we fit a ZIPSCD model  with GAR(1) intensity processes by using the filtering and estimation algorithm described in Section 4. To find the initial values to start the algorithm, we used the following factorial moment equations, namely $\bar y^{(1)}  = E\left( {Y_t } \right) = (1 - \omega)\frac{p}{\beta },\ \ \ \
\bar y^{(2)}  = E\left( {Y_t (Y_t  - 1)} \right) = (1 - \omega)\frac{{p(p + 1)}}{{\beta ^2 }} $ and
\[
\bar y^{(3)}  = E\left( {Y_t (Y_t  - 1)(Y_t  - 2)} \right) = (1 - \omega)\frac{{p(p + 1)(p + 2)}}{{\beta ^3 }}.
\]
Once the initial estimates are obtained, we run the filtering algorithm and then use these filtered intensities to compute the estimates. This process is repeated until convergence.  The final estimates of model parameters are given in Table 6. \\ 
 \begin{center} 
 	\framebox{  \textbf{Table 6  about here}.}
 \end{center}
 \vspace{0.2in}
%
%

Following  Zeger(1988),  we  have calculated the standard errors of the estimators by means of  simulations. That is, we have simulated a sample of size $n=209$ from a ZIP-GAR(1) model with the final estimates  for the data generating process and computed the new estimates. Then we repeated the  procedure 1000 times and the empirical standard deviation of these 1000 estimates is taken as the standard error of the estimates. The histogram along with a superimposed gamma density and the  plot of filtered intensities are exhibited in Figure 3. \\
\begin{center} 
	\framebox{  \textbf{Figure 3  about here}.}
\end{center}
\vspace{0.2in}
%
  
\par To check the model adequacy, we found the  Pearson residuals defined by 
\begin{equation}\label{Presid}
	{e_t} = \frac{{{y_t} - (1 - w){\lambda _t}}}{{\sqrt {\left( {1 - w} \right)\left( {1 + w{\lambda _t}} \right){\lambda _t}} }};t = 1,2,...,n
\end{equation} 
with filtered values of $\lambda_t$ and the estimate of $\omega$. The time series plot, acf and pacf of the residulas are given in Figure 4. The residulas show no significant autocorrelation at all lags considered. Further, the Ljung-Box test applied to the residual series at lag 20 confirms this fact with a $p$-value 0.4852. Finally we computed the theorotical probabilities \[
P\left( {Y_t  = k} \right) = \left\{ {\begin{array}{*{20}c}
		{\omega  + (1 - \omega )\left( {\frac{\beta }{{\beta  + 1}}} \right)^p ;\,\;\quad k = 0}  \\
		{(1 - \omega )\frac{{\beta ^p \,\Gamma \left( {p + k} \right)}}{{k!\left( {\beta  + 1} \right)^{p + k} \Gamma \left( p \right)}}\,;k \ge 1}  \\
\end{array}} \right.
\]
and compared it with empirical probabilities. Figure 5 depicts the strength of agreement between the fitted  and empirical probabilities. For instance, the sample  proportion of zeros is 0.2823 whereas  the estimated probability using the ZIPSCD-GAR(1) model is 0.2882. This shows a reasonably good fit. \\
 \begin{center} 
 	\framebox{  \textbf{Figure 4  about here}.}
 \end{center}
 \vspace{0.2in}

%
\begin{center} 
	\framebox{  \textbf{Figure 5  about here}.}
\end{center}
\vspace{0.1in}

As an application of the model to a zero deflated situation, we consider the monthly count of aggravated assaults reported in the 34th police car beat in Pittsburgh which was analyzed previously by Barreto-Souza (2015) and Sharafi et al (2020). The data is collected during the period January 1990 to December 2001. Figure 6 exhibits the time series plot, acf and histogram of the data. {\it{The data is available in Appendix C of the supporting information provided by Barreto-Souza (2015)}}.

\begin{center} 
	\framebox{  \textbf{Figure 6  about here}.}
\end{center}
\vspace{0.1in}
%
When a Poisson conditional distribution is fitted to this data we get a deflated number of zeros compared with the empirical count. This fact leads us to fit a zero deflated model using the ZMPSCD model with GAR(1) intensities. The resulting estimators are $\hat w =-0.1161, \hat \rho=0.4311, \hat p=1.8314$ and $\hat \beta=2.2575$. The negative sign of the estimate of $w,$ the zero modification parameter,  clearly indicates the presence of zero deflation compared to the base line Poisson model. The fitted probabilities are displayed in the Figure 7. 

\begin{center} 
	\framebox{  \textbf{Figure 7  about here}.}
\end{center}
\vspace{0.1in}
%
 This shows a close agreement of correct zero probability with the proposed model. The Figure 8
  shows the filtered values of latent intensities $\lambda_t$ along with actual data and the acf of Pearson residuals obtained as in (\ref{Presid}).  
The acf plot shows no remaining autocorrelations in the residuals. The  Ljung-Box test for randomness applied to the residuals confirms the absence of significant autocorrelation upto lag order 20 at $5\%$ level of significance. 

\begin{center} 
	\framebox{  \textbf{Figure 8  about here}.}
\end{center}
\vspace{0.1in}
%
%

\section{Concluding Remarks}
Count time series with excess or deficient  zeros occur in some practical situations and we proposed zero modified count time series  with  Markov dependent intensities to analyze such situations. 
It is demonstrated that the method of estimating function performs well in  filtering the unobserved intensities and then estimating  model parameters.  Yang et al (2015) introduced  a   state space model for zero inflated count series and used a Monte Carlo expectation maximization algorithm for analysis. 
As an alternative, we recommend the EF based filtering and estimation procedure. Once suitable initial estimates were obtained, the EF method provides feasible estimates for both Poisson and negative binomial models.  
We plan to develop, in the near future, coherent forecasting of zero modified count data with Markovian latent intensities. 
\vs .2cm
{\bf Acknowledgement:}  N. Balakrishna acknowledges the financial support by  Science and Engineering Research Board (SERB) of India under MATRICS scheme MTR/2018/000195. The research of Bovas Abraham was supported by a grant from the Natural Sciences and Engineering Research Council  (NSERC) of Canada. This work was initiated while Balakrishna spent a part of his sabbatical at University of Waterloo during September to November, 2019.
\vs .3in

\noi{ \large \bf References}
\bd
\itemsep -.06cm
%
%

\item Abraham, B. and Balakrishna, N. (2012). Product autoregressive models for non-negative variables. {\it Statistics and Probability Letters,} 82, 1530 - 1537.

\item Barreto-Souza, W. (2015). Zero-modified geometric INAR(1) process for modelling count time series with deflation or inflation of zeros. {\it Journal of Time Series Analysis.} 36(6), 839-52.
%
\item Bertoli, W., Katiane S. Conceição, Marinho G. Andrade and
Francisco Louzada (2019). Bayesian approach for the zero-modified Poisson–Lindley regression model. {\it Brazilian Journal of Probability and Statistics,} 33, No. 4, 826–860.

\item Bhogal SK, Thekke Variyam R.(2019). Conditional duration models for high-frequency data: a review on recent developments. {\it Journal of Economic Survey,} 33(1):252-273.

\item Cox, D. R. (1981). Statistical analysis of time series: Some recent developments. {\it Scandinavian Journal of Statistics,} 8:93-115. 


\item Davis, R. A., Holan, S. H., Lund, R. and Ravishanker, N. (2016). Handbook of discrete-valued time series. Chapman and Hall/ CRC, Boca Raton, FL.

\item Engle, RF \& Russell, JR (1998). Autoregressive conditional duration: a new approach for irregularly spaced transaction data. {\it  Journal of Econometrics,} 119, 381-482.

\item Dietz, E., Bohning, D. (2000). On estimation of the Poisson parameter in zero-modified Poisson models. {\it Computational Statistics and Data Analysis,} 34:441-459.

\item Ferland, R., Latour, A., and Oraichi, D. (2006). Integer-valued GARCH model. {\it Journal of Time Series Analysis,} 27:923-942.

 \item Fokianos, K. (2016). Statistical analysis of count time series models: A GLM perspective.  Handbook of discrete-valued time series. Chapman and Hall/ CRC, Boca Raton, FL. 
   
 \item Gaver, D. P., Lewis, P. A. W. (1980).
{First order autoregressive gamma sequences and point processes.}
 { \it Advances in Applied Probability,} 12: 727-745.  
    
\item Godambe, V. P. (1985). 
{The Foundations of Finite Sample Estimation in Stochastic Processes.}
 { \it Biometrika,} 72: 419-428.    

\item Lambert, D.  (1992).  Zero-inflated Poisson Regression, With an Application to Defects in
Manufacturing.  {\it{Technometrics,}} 34, 1–14.

\item Lawrance A. J. (1982). The innovation distribution
of a gamma distributed autoregressive process.
{\it Scandinavian Journal of Statistics} , 9, 224-236.

\item Lawrance, A. J. and Lewis, P. A. W. (1981). A new autoregressive time series model in exponential variables (NEAR(1)). {\it Advances in Applied Probability} 13,  826 - 845.  

 \item  Lewis, P. A .W.,  McKenzie, E. \& Hugus, D. K. (1989). Gamma processes,
 {\it Stochastic Models,} 5:1, 1-30.
%
%
 \item Muhammed Anvar, P., Balakrishna, N. and Bovas Abraham (2019). Stochastic volatility generated by product autoregressive
 models. {\it Stat.,DOI: 10.1002/sta4.232}
 
 \item Naik-Nimbalkar, U. V. and Rajarshi, M. B. (1995). Filtering and Smoothing Via Estimating Functions, {\it Journal of American Statistical Association,} 90, 301-306 .

 \item Pitt, M.K., Walker, S.G. (2005). Constructing stationary time series models using auxiliary variables with applications. {\it Journal of American statistical association,} 100, 554-564.

\item  Thekke, R., Mishra, A., and  Abraham, B. (2016). Estimation, filtering and smoothing in the stochastic conditional duration model: an estimating function approach. {\it Stat}, 5(1), 11-21.

\item Sharafi, M., Sajjadnia, Z. and Zamani, A. (2021): A first-order integer-valued autoregressive process with zero-modified Poisson-Lindley distributed innovations, {\it{Communications in Statistics - Simulation and Computation,}} DOI:10.1080/03610918.2020.1864644

%
%
\item Sim, C. H. (1990). First order autoregressive models for gamma
and exponential processes, {\it Journal of Applied Probability,} 27, 325-332.

\item Tjostheim, D. (2016). Count time series with observation-driven autoregressive parameter dynamics.  Handbook of discrete-valued time series. Chapman and Hall/ CRC.pp 77-100.
 \item Thavaneswaran, A., Ravishanker, N., Liang, Y. (2015).
 {Generalized duration models and optimal estimation using estimating functions.} {\it Annals of the Institute of Statistical Mathematics,} 67(1):129-156.
 
  \item Yang, M., Cavanaugh, J. E., Zamba, G. K.D. (2015). State space models for count time series with excess zeros. {\it Statistical Modelling,} 15(1), 70-90.
 
 \item Zehnwirth, B (1988). A generalization of the Kalman filter model for models with state-dependent observation variance. {\it Journal of the American Statistical Association,} 83, 164-167.
 
  \item Zeger, S. L. (1988). A regression model for time series of counts. {\it Biometrika}, 75(4), 621-629.
 
 \item Zhu, F. (2012). Zero-inflated Poisson and negative binomial integer-valued GARCH models. {\it Journal
 of Statistical Planning and Inference,} 142:826-839.
\ed

\newpage

\begin{table}
	\begin{center}	
		\caption{ Estimation results of ZIPSCD with GAR(1) intensities. Estimates of paramters along with their MSE are given in parenthesis.}
		\begin{tabular}{|c|c|c|c|c|c|c|c|c|}
			\hline
			\multirow{2}[4]{*}{\textbf{$n$}} & \multicolumn{4}{c|}{True values} & \multicolumn{4}{c|}{Estimates} \\
			\cline{2-9}          & $\rho$   & $\omega$     & $\beta$ & $p$     & \multicolumn{1}{c|}{$\hat\rho$} & \multicolumn{1}{c|}{$\hat{\omega}$} & \multicolumn{1}{c|}{$\hat{\beta}$} & \multicolumn{1}{c|}{$\hat{p}$ } \\
			\hline
			\multirow{10}[10]{*}{200} & 
			\multirow{2}[2]{*}{0.95} & \multirow{2}[2]{*}{0.3} & \multirow{2}[2]{*}{0.5} & \multirow{2}[2]{*}{2} & 0.9398
			& 0.3030
			& 0.5928
			& 2.5129
			\\
			&       &       &       &       & (0.0003
			) & (0.0007
			) & (0.0425
			) & (0.4123
			)  \\
			\cline{2-9}          & \multirow{2}[2]{*}{0.9} & \multirow{2}[2]{*}{0.2} & \multirow{2}[2]{*}{1.5} & \multirow{2}[2]{*}{3} & 0.8911
			&0.2059
			& 1.6258
			& 3.2966
			\\
			&       &       &       &       & (0.0004
			) & (0.0022
			) & (0.0879) & (0.3917
			
			)   \\
			\cline{2-9}          & \multirow{2}[2]{*}{0.8} & \multirow{2}[2]{*}{0.2} & \multirow{2}[2]{*}{2} & \multirow{2}[2]{*}{4} &0.7924
			& 0.2012
			& 2.1731
			& 4.1931
			\\
			&       &       &       &       & (0.0007) & (0.0024
			) & (0.08245
			) & (0.39331
			)  \\
			\cline{2-9}          & \multirow{2}[2]{*}{0.6} & \multirow{2}[2]{*}{0.2} & \multirow{2}[2]{*}{1.5} & \multirow{2}[2]{*}{1.5} & 0.5922
			& 0.2068
			& 1.4682
			& 1.4581
			\\
			&       &       &       &       & (0.0012
			) & (0.0051
			) & (0.0599
			) & (0.0591
			)  \\
			\cline{2-9}          & \multirow{2}[2]{*}{0.5} & \multirow{2}[2]{*}{0.4} & \multirow{2}[2]{*}{0.75} & \multirow{2}[2]{*}{3.5} & 0.4981
			& 0.3993
			& 0.8318
			& 3.8705
			\\
			&       &       &       &       & (0.0014
			) & (0.0005
			) & (0.0127
			) & (0.0967
			)  \\

			\hline
			\multirow{10}[10]{*}{1000} & \multirow{2}[2]{*}{0.95} & \multirow{2}[2]{*}{0.3} & \multirow{2}[2]{*}{0.5} & \multirow{2}[2]{*}{2} & 0.9441 & 0.3016 & 0.5727 & 2.2257  \\
			&       &       &       &       & (0.0001) & (0.0001) & (0.0257) & (0.368)  \\
			\cline{2-9}          & \multirow{2}[2]{*}{0.9} & \multirow{2}[2]{*}{0.2} & \multirow{2}[2]{*}{1.5} & \multirow{2}[2]{*}{3} & 0.8939 & 0.2028 & 1.5932 & 3.1523  \\
			&       &       &       &       & (0.0002) & (0.0008) & (0.0814) & (0.3108)  \\
			\cline{2-9}          & \multirow{2}[2]{*}{0.8} & \multirow{2}[2]{*}{0.2} & \multirow{2}[2]{*}{2} & \multirow{2}[2]{*}{4} & 0.797249 & 0.2009 & 2.0293 & 4.0342  \\
			&       &       &       &       & (0.0002) & (0.0004) & (0.0326) & (0.1331)  \\
			\cline{2-9}          & \multirow{2}[2]{*}{0.6} & \multirow{2}[2]{*}{0.2} & \multirow{2}[2]{*}{1.5} & \multirow{2}[2]{*}{1.5} & 0.5964 & 0.2004 & 1.5179 & 1.5164  \\
			&       &       &       &       & (0.0004) & (0.0011) & (0.0127) & (0.0124)  \\
			\cline{2-9}          & \multirow{2}[2]{*}{0.5} & \multirow{2}[2]{*}{0.4} & \multirow{2}[2]{*}{0.75} & \multirow{2}[2]{*}{3.5} & 0.4975 & 0.4006 & 0.7541 & 3.5117 \\
			&       &       &       &       & (0.0005) & (0.0006) & (0.0018) & (0.0375)  \\
			\hline
		\end{tabular}%
		\label{table1}%
	\end{center}
\end{table}%

\newpage 

\begin{table}[h!]
	\begin{center}
		\caption{ Estimation results of ZINBSCD with GAR(1) intensities. Estimates of paramters along with their MSE in parenthesis are given.}
		
		\begin{tabular}{|c|c|c|c|c|c|c|c|c|c|c|}
			\hline
			\multirow{2}[4]{*}{$n$} & \multicolumn{5}{c|}{True values}      & \multicolumn{5}{c|}{Estimates} \\
			\cline{2-11}          & $\rho$   & $\omega$     & $\beta$ & $p$     & $a$     & \multicolumn{1}{c|}{$\hat\rho$} & \multicolumn{1}{c|}{$\hat{\omega}$} & \multicolumn{1}{c|}{$\hat{\beta}$ } & \multicolumn{1}{c|}{$\hat p$} & \multicolumn{1}{c|}{$\hat a$} \\
			\hline
			\multirow{10}[10]{*}{200} & \multirow{2}[2]{*}{0.95} & \multirow{2}[2]{*}{0.3} & \multirow{2}[2]{*}{1.5} & \multirow{2}[2]{*}{0.75} & \multirow{2}[2]{*}{1.5} & 0.8994
			& 0.3504
			& 1.4237
			& 0.4675
			& 1.7922
			\\
			&       &       &       &       &       & (0.0035
			) & (0.0065
			) & (0.0686
			) & (0.0813
			) & (0.4036) \\
			\cline{2-11}          & \multirow{2}[2]{*}{0.9} & \multirow{2}[2]{*}{0.2} & \multirow{2}[2]{*}{2} & \multirow{2}[2]{*}{3} & \multirow{2}[2]{*}{0.5} & 0.8568
			& 0.2236
			& 2.4111
			& 3.8124
			& 0.3727
			\\
			&       &       &       &       &       & (0.0063
			) & (0.0080
			) & (0.0310
			) & (0.0880
			) & (0.3948
			) \\
			\cline{2-11}          & \multirow{2}[2]{*}{0.8} & \multirow{2}[2]{*}{0.3} & \multirow{2}[2]{*}{2} & \multirow{2}[2]{*}{1} & \multirow{2}[2]{*}{0.5} & 0.7534
			& 0.3518
			& 2.05311
			& 1.10356
			& 0.3929
			\\
			&       &       &       &       &       & (0.0036
			) & (0.0135
			) & (0.0059
			) & (0.0352
			) & (0.0341
			) \\
			\cline{2-11}          & \multirow{2}[2]{*}{0.6} & \multirow{2}[2]{*}{0.2} & \multirow{2}[2]{*}{3} & \multirow{2}[2]{*}{4.5} & \multirow{2}[2]{*}{0.75} & 0.5791
			& 0.2497
			& 2.8082
			& 4.2139
			& 0.5686
			\\
			&       &       &       &       &       & (0.0028
			) & (0.0058
			) & (0.0522
			) & (0.9531
			) & (0.0914
			) \\
			\cline{2-11}          & \multirow{2}[2]{*}{0.5} & \multirow{2}[2]{*}{0.3} & \multirow{2}[2]{*}{2} & \multirow{2}[2]{*}{1.5} & \multirow{2}[2]{*}{0.25} & 0.4741
			& 0.3105
			& 2.2265
			& 1.8927
			& 0.1154
			\\
			&       &       &       &       &       & (0.0030
			) & (0.0053
			) & (0.00902
			) & (0.3684
			) & (0.1451
			) \\
			\hline
			
			\multirow{10}[10]{*}{1000} & \multirow{2}[2]{*}{0.95} & \multirow{2}[2]{*}{0.3} & \multirow{2}[2]{*}{1.5} & \multirow{2}[2]{*}{0.75} & \multirow{2}[2]{*}{1.5} & 0.9314 & 0.3201 & 1.5338 & 0.5787 & 1.5551 \\
			&       &       &       &       &       & (0.0004) & (0.0026) & (0.0126) & (0.0391) & (0.1898) \\
			\cline{2-11}          & \multirow{2}[2]{*}{0.9} & \multirow{2}[2]{*}{0.2} & \multirow{2}[2]{*}{2} & \multirow{2}[2]{*}{3} & \multirow{2}[2]{*}{0.5} & 0.8599 &0.2579 & 2.1525 & 3.4919 & 0.3692 \\
			&       &       &       &       &       & (0.0018) & (0.0052) & (0.0570) & (0.5613) & (0.0411) \\
			\cline{2-11}          & \multirow{2}[2]{*}{0.8} & \multirow{2}[2]{*}{0.3} & \multirow{2}[2]{*}{2} & \multirow{2}[2]{*}{1} & \multirow{2}[2]{*}{0.5} & 0.7812 & 0.3041 & 2.0097 & 1.0754 & 0.463 \\
			&       &       &       &       &       & (0.0007) & (0.0008) & (0.0026) & (0.0102) & (0.0128) \\
			\cline{2-11}          & \multirow{2}[2]{*}{0.6} & \multirow{2}[2]{*}{0.2} & \multirow{2}[2]{*}{3} & \multirow{2}[2]{*}{4.5} & \multirow{2}[2]{*}{0.75} & 0.5618 & 0.2181 & 3.0776 & 4.3727 & 0.5207 \\
			&       &       &       &       &       & (0.0019) & (0.0012) & (0.0145) & (0.7308) & (0.0697) \\
			\cline{2-11}          & \multirow{2}[2]{*}{0.5} & \multirow{2}[2]{*}{0.3} & \multirow{2}[2]{*}{2} & \multirow{2}[2]{*}{1.5} & \multirow{2}[2]{*}{0.25} & 0.4791 & 0.3029 & 2.0078 & 1.8631 & 0.1358 \\
			&       &       &       &       &       & (0.0011) & (0.0006) & (0.0011) & (0.2910) & (0.0172) \\
			\hline
		\end{tabular}%
		\label{tab:addlabel}%
	\end{center}
\end{table}%

\newpage 
 \begin{table}[h!]
 	\centering
 	\caption{Estimation results of ZDPSCD with GAR(1) intensities. Average estimates of parameters along with their MSE in parenthesis  are given.}
 	\begin{tabular}{|c|c|c|c|c|c|c|c|c|}
 		
 		\hline
 		\multirow{2}[4]{*}{$n$} & \multicolumn{4}{c|}{True values} & \multicolumn{4}{c|}{Estimates} \\
 		\cline{2-9}          & $\rho$ & $\omega$ & $\b$ & $p$ & \multicolumn{1}{c|}{$\hat \rho$} & \multicolumn{1}{c|}{$\hat \omega$} & \multicolumn{1}{c|}{$\hat \b$} & \multicolumn{1}{c|}{$\hat p$} \\
 		\hline
 		\multirow{10}[10]{*}{200} & \multirow{2}[2]{*}{0.95} & \multirow{2}[2]{*}{-0.3} & \multirow{2}[2]{*}{0.5} & \multirow{2}[2]{*}{2} & 0.9373 & -0.3269 & 0.5979 & 1.8253 \\
 		&       &       &       &       & (0.0003) & (0.0063) & (0.4759) & (0.8807) \\
 		\cline{2-9}          & \multirow{2}[2]{*}{0.9} & \multirow{2}[2]{*}{-0.2} & \multirow{2}[2]{*}{1.5} & \multirow{2}[2]{*}{3} & 0.8901 & -0.2141 & 1.5138 & 3.9451 \\
 		&       &       &       &       & (0.0004) & (0.0421) & (0.0067) & (0.9821) \\
 		\cline{2-9}          & \multirow{2}[2]{*}{0.8} & \multirow{2}[2]{*}{-0.2} & \multirow{2}[2]{*}{0.5} & \multirow{2}[2]{*}{4} & 0.7936 & -0.2206 & 0.5274 & 4.9454 \\
 		&       &       &       &       & (0.0006) & (0.0062) & (0.1454) & (0.8745) \\
 		\cline{2-9}          & \multirow{2}[2]{*}{0.6} & \multirow{2}[2]{*}{-0.2} & \multirow{2}[2]{*}{0.75} & \multirow{2}[2]{*}{1.5} & 0.5924 & -0.2005 & 0.7502 & 1.6945 \\
 		&       &       &       &       & (0.0012) & (0.0089) & (0.0158) & (0.9172) \\
 		\cline{2-9}          & \multirow{2}[2]{*}{0.5} & \multirow{2}[2]{*}{-0.1} & \multirow{2}[2]{*}{0.75} & \multirow{2}[2]{*}{3.5} & 0.4891 & -0.1177 & 0.7556 & 3.7843 \\
 		&       &       &       &       & (0.0015) & (0.0235) & (0.0076) & (0.8325) \\
 		\hline
 		\multirow{10}[10]{*}{1000} & \multirow{2}[2]{*}{0.95} & \multirow{2}[2]{*}{-0.3} & \multirow{2}[2]{*}{0.5} & \multirow{2}[2]{*}{2} & 0.9461 & -0.3058 & 0.5215 & 1.4814 \\
 		&       &       &       &       & (0.0001) & (0.0008) & (0.1321) & (0.73056) \\
 		\cline{2-9}          & \multirow{2}[2]{*}{0.9} & \multirow{2}[2]{*}{-0.2} & \multirow{2}[2]{*}{1.5} & \multirow{2}[2]{*}{3} & 0.8965 & -0.2025 & 1.5015 & 3.4686 \\
 		&       &       &       &       & (0.0001) & (0.0102)	 & (0.0011) & (0.2536) \\
 		\cline{2-9}          & \multirow{2}[2]{*}{0.8} & \multirow{2}[2]{*}{-0.2} & \multirow{2}[2]{*}{0.5} & \multirow{2}[2]{*}{4} & 0.7971 & -0.2013 & 0.5019 & 4.1444 \\
 		&       &       &       &       & (0.0002) & (0.0008) & (0.0255) & (0.1764) \\
 		\cline{2-9}          & \multirow{2}[2]{*}{0.6} & \multirow{2}[2]{*}{-0.2} & \multirow{2}[2]{*}{0.75} & \multirow{2}[2]{*}{1.5} & 0.5974 & -0.1957 & 0.7483 & 1.5852 \\
 		&       &       &       &       & (0.0005) & (0.0016) & (0.0025) & (0.5028) \\
 		\cline{2-9}          & \multirow{2}[2]{*}{0.5} & \multirow{2}[2]{*}{-0.1} & \multirow{2}[2]{*}{0.75} & \multirow{2}[2]{*}{3.5} & 0.4945 & -0.1025 & 0.7507 & 3.5835 \\
 		&       &       &       &       & (0.0005) & (0.0048) & (0.0015) & (0.2309) \\
 		\hline
 	\end{tabular}%
 	\label{Table3}%
 \end{table}%
 
 \newpage
 \begin{table}[h!]
 	\centering
 	\caption{Estimation results of ZDNBSCD model with GAR(1) intensities. Average estimates of paramters along with their MSE in parenthesis are  given. }
 	\begin{tabular}{|c|c|c|c|c|c|c|c|c|c|c|}
 		\hline
 		\multirow{2}[4]{*}{$n$} & \multicolumn{5}{c|}{True values} & \multicolumn{5}{c|}{Estimates} \\
 		\cline{2-11}          & $\rho$ & $\omega$ & $\b$ & $p$ & $a$ & \multicolumn{1}{c|}{$\hat \rho$} & \multicolumn{1}{c|}{$\hat \omega$} & \multicolumn{1}{c|}{$\hat \b$} & \multicolumn{1}{c|}{$\hat p$} & \multicolumn{1}{c|}{$\hat a$} \\
 		\hline
 		\multirow{10}[10]{*}{200} & \multirow{2}[2]{*}{0.95} & \multirow{2}[2]{*}{-0.3} & \multirow{2}[2]{*}{0.5} & \multirow{2}[2]{*}{2} & \multirow{2}[2]{*}{1.5} & 0.9055 & -0.236 & 0.5755 & 2.3650 & 1.3004 \\
 		&       &       &       &       &       & (0.0031) & (0.0216) & (0.9042) & (0.2077) & (0.2239) \\
 		\cline{2-11}          & \multirow{2}[2]{*}{0.9} & \multirow{2}[2]{*}{-0.2} & \multirow{2}[2]{*}{1.5} & \multirow{2}[2]{*}{3} & \multirow{2}[2]{*}{0.5} & 0.8601 & -0.1814 & 1.237 & 2.5047 & 0.3724 \\
 		&       &       &       &       &       & (0.0031) & (0.0047) & (0.0586) & (0.2272) & (0.0461) \\
 		\cline{2-11}          & \multirow{2}[2]{*}{0.8} & \multirow{2}[2]{*}{-0.2} & \multirow{2}[2]{*}{0.5} & \multirow{2}[2]{*}{2} & \multirow{2}[2]{*}{0.25} & 0.7373 & -0.1707 & 0.3942 & 1.5529 & 0.1505 \\
 		&       &       &       &       &       & (0.0064) & (0.0022) & (0.2482) & (0.2072) & (0.0161) \\
 		\cline{2-11}          & \multirow{2}[2]{*}{0.6} & \multirow{2}[2]{*}{-0.2} & \multirow{2}[2]{*}{1.5} & \multirow{2}[2]{*}{1.5} & \multirow{2}[2]{*}{0.75} & 0.5601 & -0.1711 & 0.9421 & 0.9621 & 0.3979 \\
 		&       &       &       &       &       & (0.0039) & (0.0086) & (0.0074) & (0.2269) & (0.2164) \\
 		\cline{2-11}          & \multirow{2}[2]{*}{0.5} & \multirow{2}[2]{*}{-0.1} & \multirow{2}[2]{*}{0.75} & \multirow{2}[2]{*}{2.5} & \multirow{2}[2]{*}{1.25} & 0.4753 & -0.0855 & 0.5263 & 1.7712 & 0.9816 \\
 		&       &       &       &       &       & (0.0031) & (0.0071) & (0.0279) & (0.9404) & (0.1896) \\
 		\hline
 		\multirow{10}[10]{*}{1000} & \multirow{2}[2]{*}{0.95} & \multirow{2}[2]{*}{-0.3} & \multirow{2}[2]{*}{0.5} & \multirow{2}[2]{*}{2} & \multirow{2}[2]{*}{1.5} & 0.9286 & -0.2522 & 0.5183 & 2.1345 & 1.4193 \\
 		&       &       &       &       &       & (0.0006) & (0.0062) & (0.1947) & (0.1134) & (0.0666) \\
 		\cline{2-11}          & \multirow{2}[2]{*}{0.9} & \multirow{2}[2]{*}{-0.2} & \multirow{2}[2]{*}{1.5} & \multirow{2}[2]{*}{3} & \multirow{2}[2]{*}{0.5} & 0.8793 & -0.1851 & 1.1773 & 2.3735 & 0.3935 \\
 		&       &       &       &       &       & (0.0006) & (0.0014) & (0.0157) & (0.1791) & (0.0181) \\
 		\cline{2-11}          & \multirow{2}[2]{*}{0.8} & \multirow{2}[2]{*}{-0.2} & \multirow{2}[2]{*}{2} & \multirow{2}[2]{*}{2} & \multirow{2}[2]{*}{0.25} & 0.7578 & -0.1753 & 0.3839 & 1.5099 & 0.1671 \\
 		&       &       &       &       &       & (0.0022) & (0.0008) & (0.0575) & (0.44448 & (0.0092) \\
 		\cline{2-11}          & \multirow{2}[2]{*}{0.6} & \multirow{2}[2]{*}{-0.2} & \multirow{2}[2]{*}{1.5} & \multirow{2}[2]{*}{1.5} & \multirow{2}[2]{*}{0.75} & 0.5669 & -0.1734 & 0.9616 & 0.9802 & 0.4003 \\
 		&       &       &       &       &       & (0.0015) & (0.0024) & (0.0017) & (0.1633) & (0.1411) \\
 		\cline{2-11}          & \multirow{2}[2]{*}{0.5} & \multirow{2}[2]{*}{-0.1} & \multirow{2}[2]{*}{0.75} & \multirow{2}[2]{*}{2.5} & \multirow{2}[2]{*}{1.25} & 0.4814 & -0.0911 & 0.5439 & 1.8283 & 1.0007 \\
 		&       &       &       &       &       & (0.0008) & (0.0016) & (0.0073) & (0.5297) & (0.0883) \\
 		\hline
 	\end{tabular}%
 	\label{tab:addlabel}%
 \end{table}%

 \newpage 
 \begin{table}[h!]
 	\caption{ Summary statistics of syphilis data.}
 	\begin{tabular}{|c|c|c|c|c|c|c|}	 	
 		\hline 
 		Sample size&Mimimum& Mean  & Median  & Maximum & No of zeros &  Variance\\ \hline
 		209&0 & 3.474 & 3 & 15 & 59 &9.2793\\  \hline
 	\end{tabular} 
 \end{table}

\begin{table}[h!]
	\begin{center}
		\caption{ Parameter estimates of syphilis data by fitting a ZIP-GAR(1) model}
		\begin{tabular}{cccc}
			\hline \hline 
			Parameter &  Estimate & S.E  \\
			\hline \hline 
			$\rho$ &  0.7492 & 0.00154\\
			\hline
			$p$ &  9.9184&  0.15874 \\
			\hline  $\beta$&  2.1275 & 0.05432 \\
			\hline 
			$\omega$&  0.2723 & 0.00936  \\
			\hline 
			
		\end{tabular} 
	\end{center}
\end{table}

\newpage	
\begin{figure}
	\centering
	\includegraphics[]{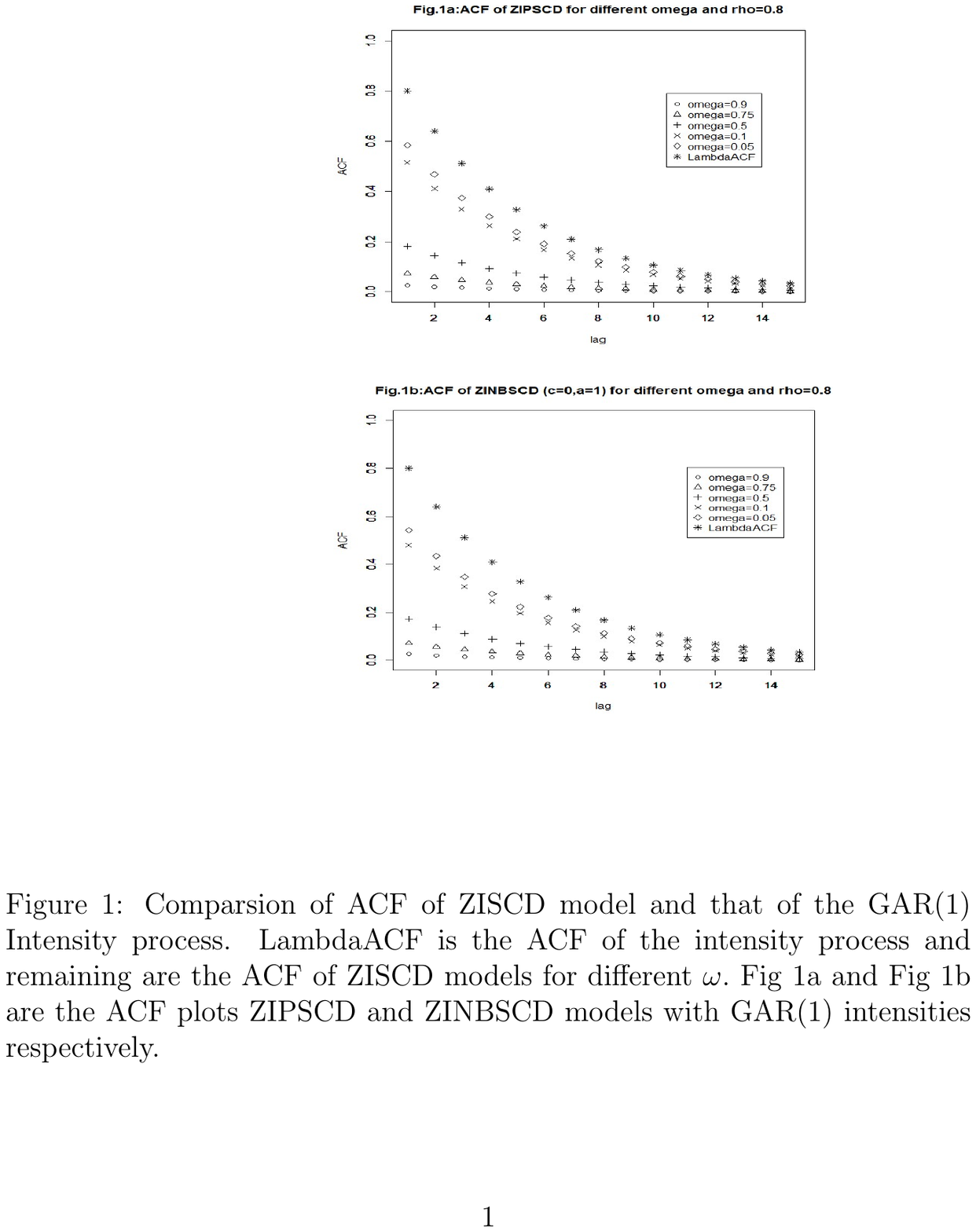}
	\caption{Figure 1: Comparsion of ACF of ZISCD model and that of the GAR(1) Intensity process. LambdaACF is the ACF of the intensity process and remaining are the ACF of ZISCD models for different $\omega.$ Fig 1a and Fig 1b are the ACF plots ZIPSCD and ZINBSCD models with GAR(1) intensities respectively. }
\end{figure}	

\begin{figure}[htbp]
	\begin{center}
	\includegraphics[]{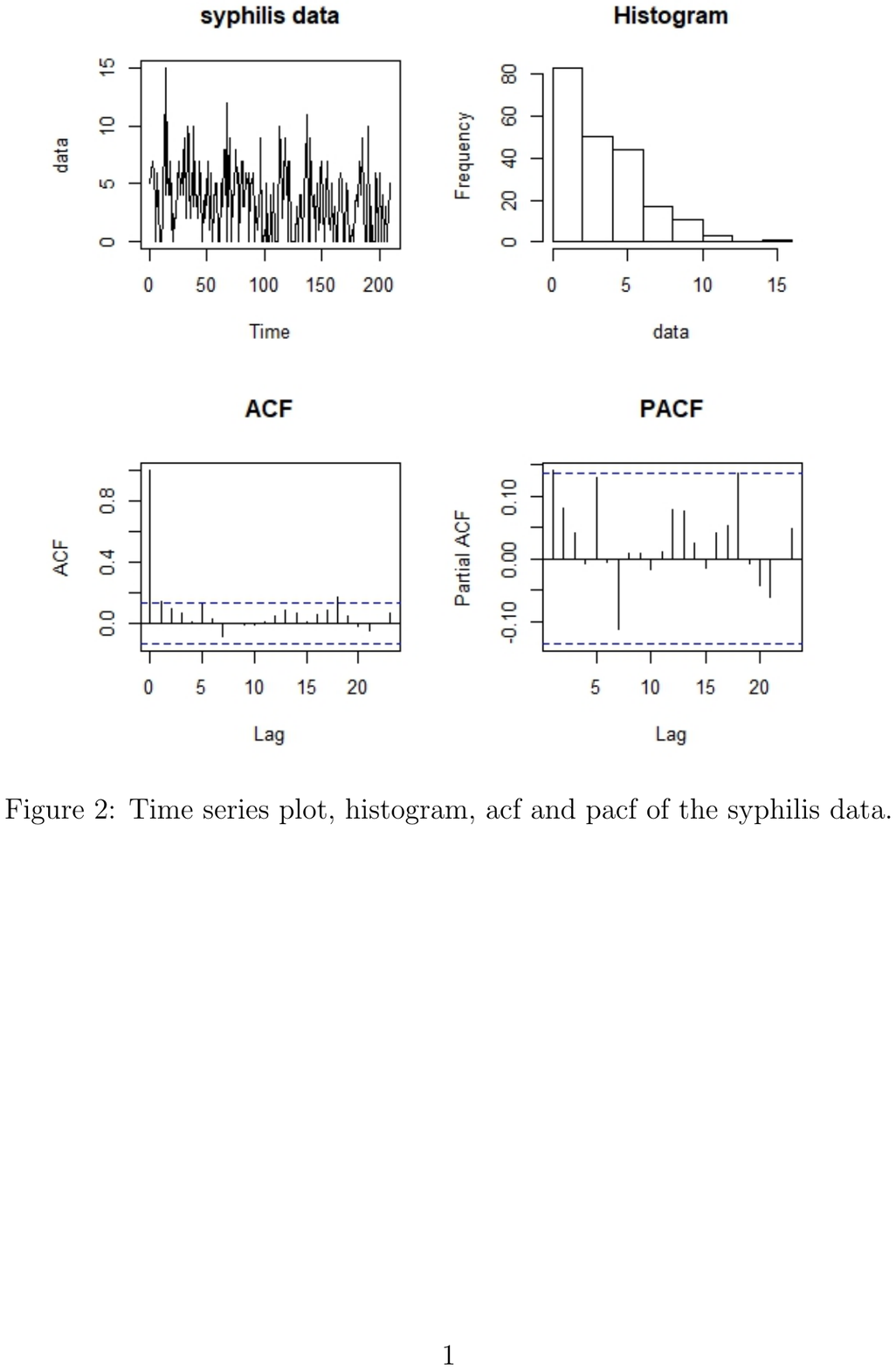}
	\caption{Figure 2: Time series plot, histogram, acf and pacf of the syphilis data.} 
	\end{center}
\end{figure}

\begin{figure}[h!]
	\centering
	\includegraphics[]{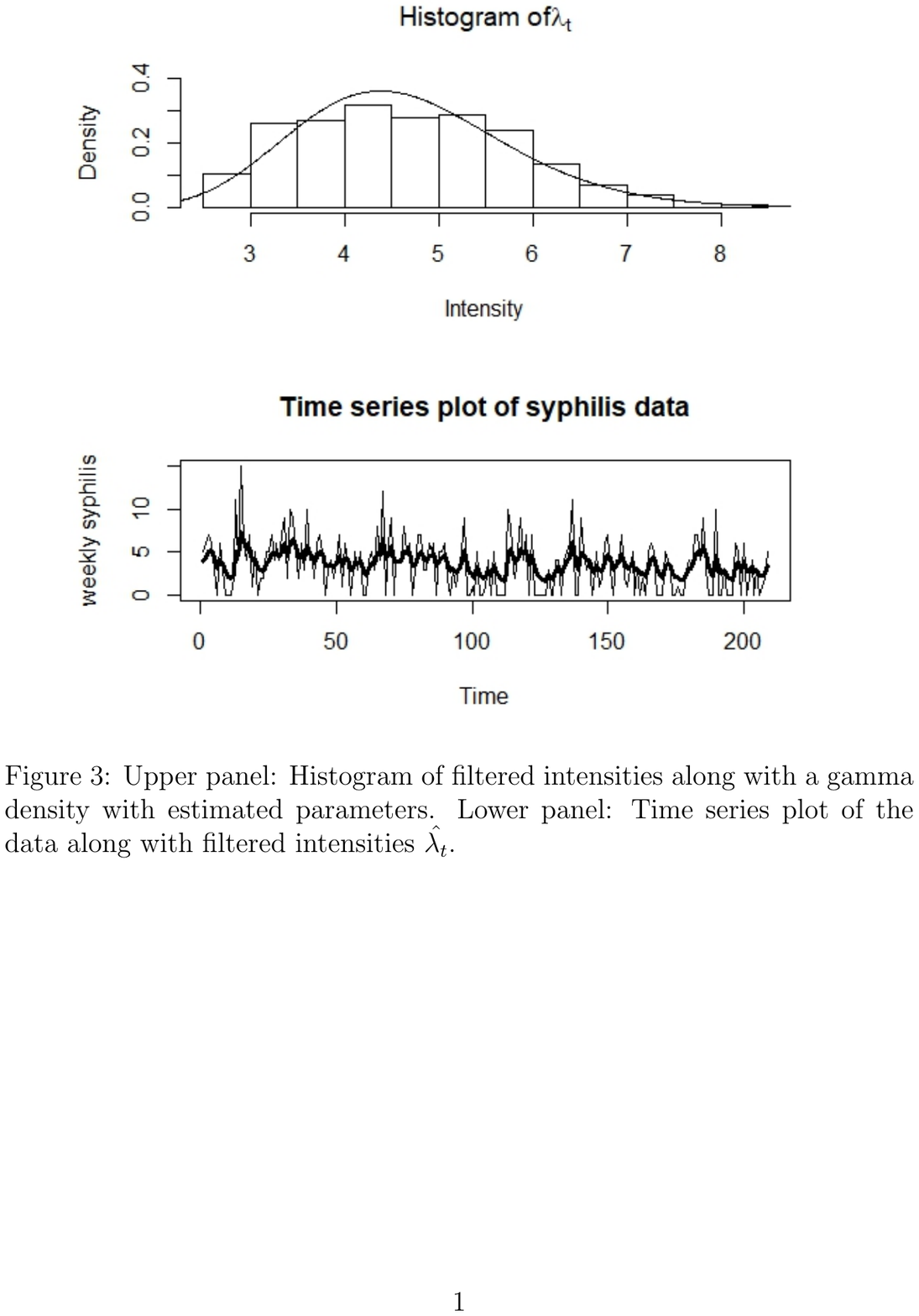}
	\caption{Figure 3: Upper panel: Histogram of filtered intensities along with a gamma density with estimated parameters. Lower panel: Time series plot of the data along with filtered intensities 
		$\hat{\lambda_t}.$}
	\label{fig:filteredintensity}
\end{figure}

\begin{figure}[h!]
	\centering
	\includegraphics[]{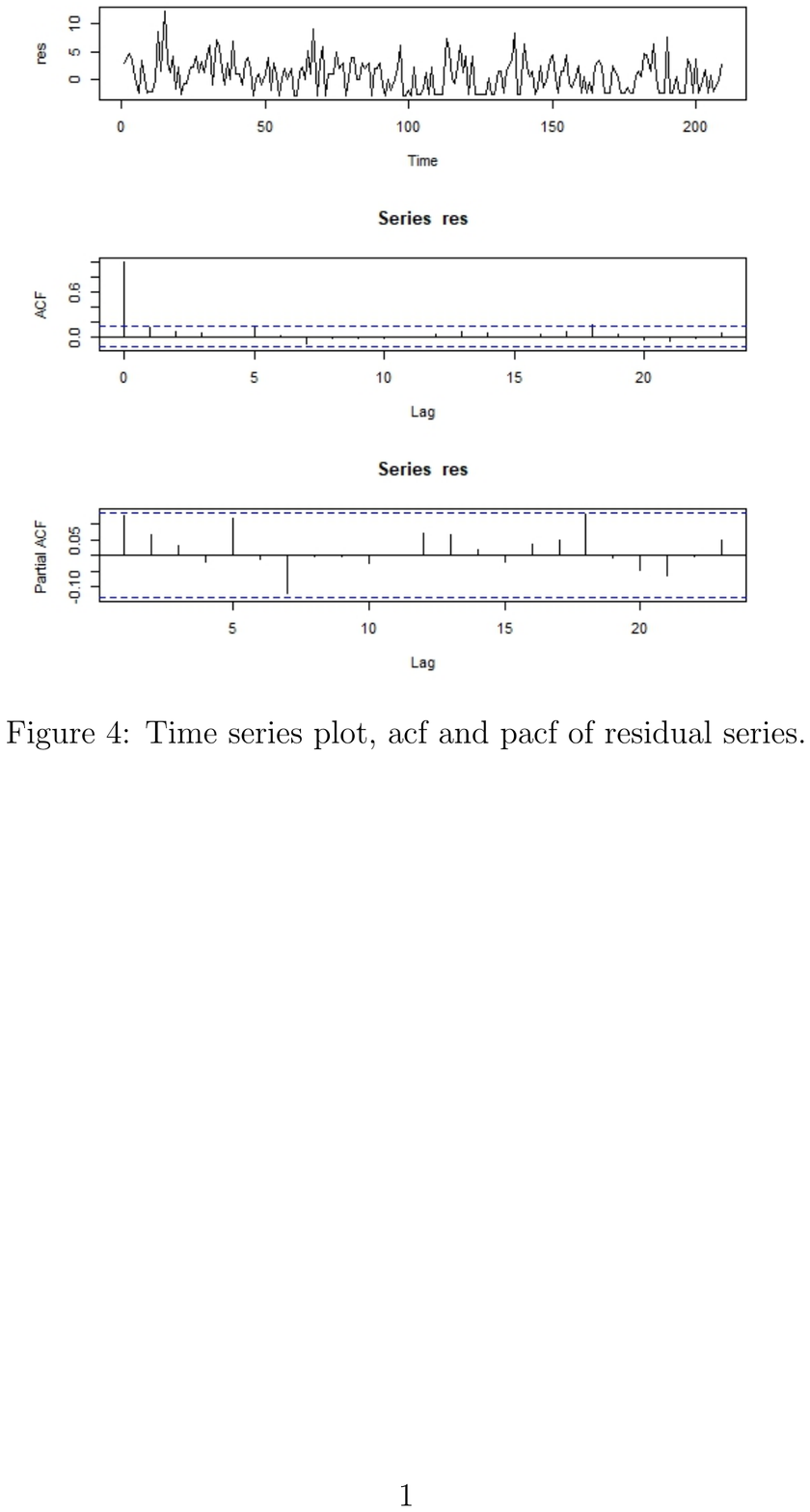}
	\caption{Figure 4: Time series plot, acf and pacf of residual series.}
	\label{fig:residualplots}
\end{figure}

\begin{figure}[h!]
	\centering
	\includegraphics[]{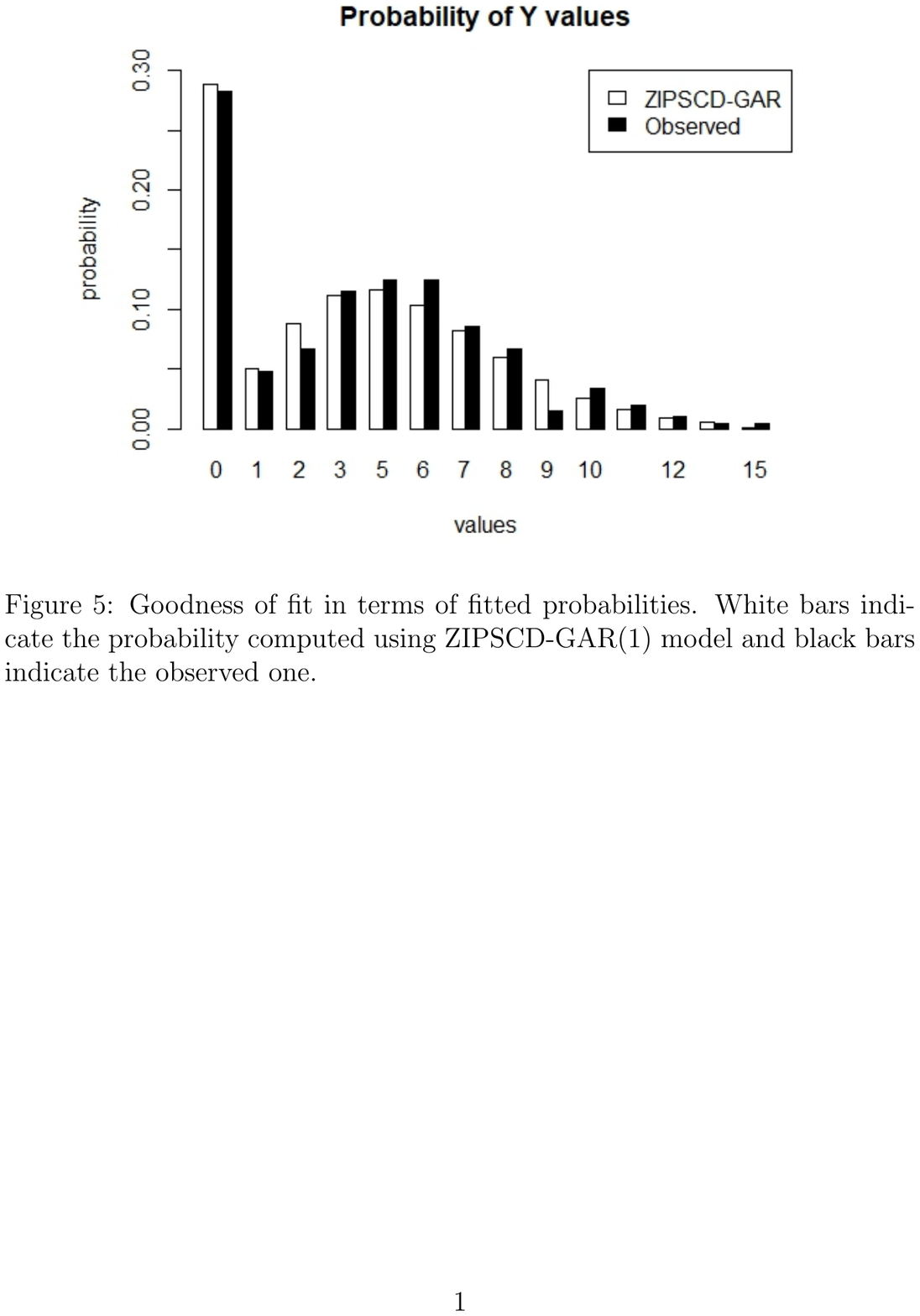}
	\caption{Figure 5: Goodness of fit in terms of fitted probabilities. White bars indicate the probability computed using ZIPSCD-GAR(1) model  and black bars indicate the observed one.}
	\label{fitted-prob-data}
\end{figure}

\begin{figure}[h!]
	\centering
	\includegraphics[]{"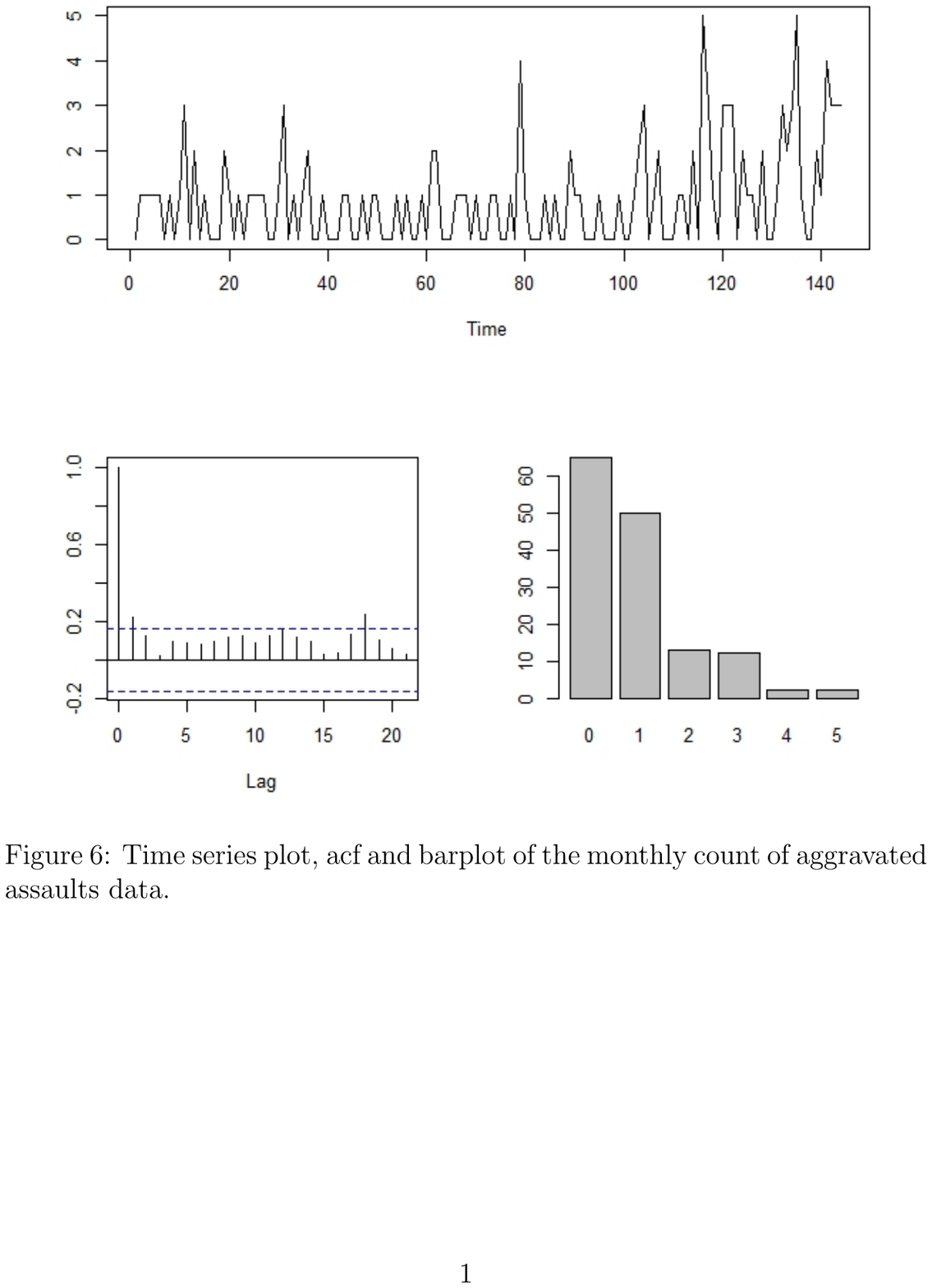}
	\caption{Figure 6: Time series plot, acf and barplot of the monthly count of aggravated assaults data.}
	\label{fig:time-series-plot-of-deflated-data}
\end{figure}

\begin{figure}[h!]
	\centering
	\includegraphics[]{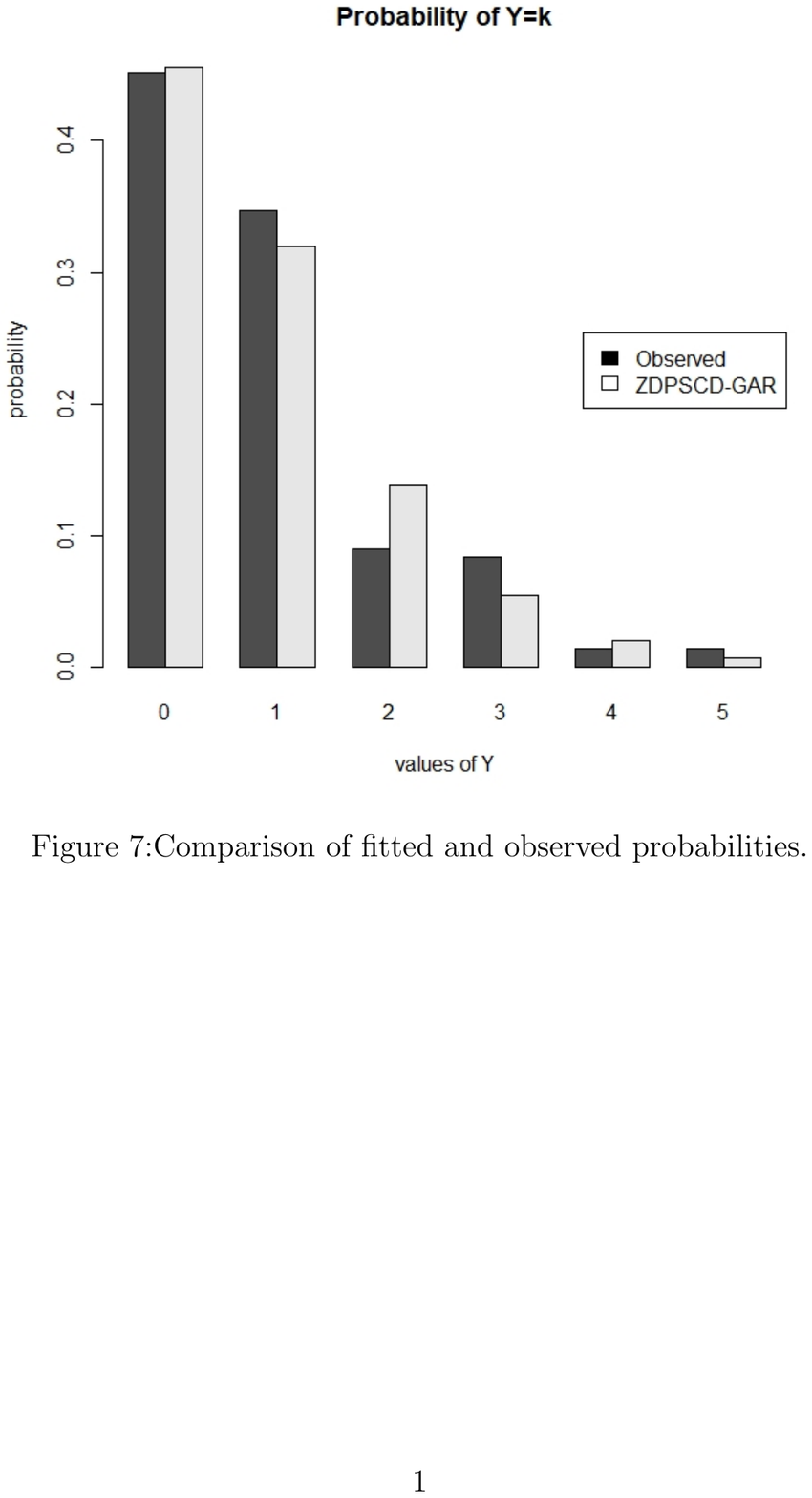}
	\caption{Figure 7:Comparison of fitted and observed probabilities. }
	\label{fittedbarplotdeflated}
\end{figure}
\begin{figure}[h!]
	\centering
	\includegraphics[]{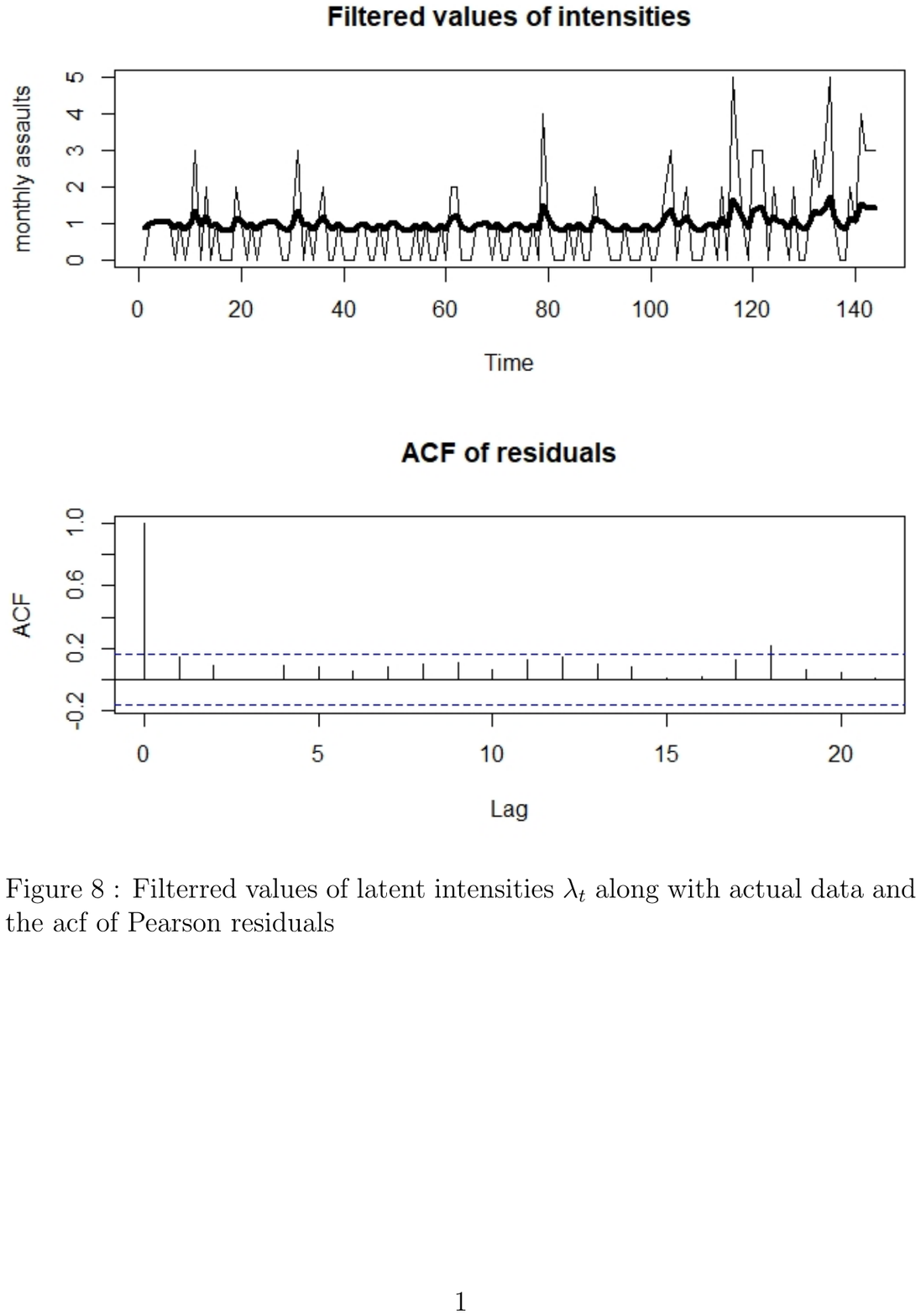}
	\caption{Figure 8 : Filterred values of latent intensities $\lambda_t$ along with actual data and the acf of Pearson residuals}
	\label{deflatedfiltered-and-residuals}
\end{figure}

\end{document}